\documentclass[1p,sort&compress,times]{elsarticle}

\usepackage{graphicx}
\usepackage{amsmath, amssymb, amscd, latexsym, bm, braket, slashed, comment}
\usepackage[colorlinks=true,pdfstartview=FitV,linkcolor=blue,citecolor=magenta,urlcolor=blue,bookmarks=true,bookmarksnumbered=true]{hyperref}

\newcommand{\bE}{{\bm E}}
\newcommand{\bB}{{\bm B}}

\begin{document}

\title{Chiral anomaly in a (1+1)-dimensional Floquet system under high-frequency electric fields}

\author[tokyo]{Kenji Fukushima}
\author[kek,sokendai,riken,qup]{Yoshimasa Hidaka}
\author[tokyo]{Takuya Shimazaki}
\author[riken]{Hidetoshi Taya}
\ead{hidetoshi.taya@riken.jp}
\address[tokyo]{Department of Physics, The University of Tokyo, 7-3-1 Hongo, Bunkyo-ku, Tokyo 113-0033, Japan}
\address[kek]{KEK Theory Center, Tsukuba 305-0801, Japan}
\address[sokendai]{Graduate University for Advanced Studies (Sokendai), Tsukuba 305-0801, Japan}
\address[riken]{RIKEN iTHEMS, RIKEN, Wako 351-0198, Japan}
\address[qup]{International Center for Quantum-field Measurement Systems for Studies of the Universe and Particles (QUP), KEK, Tsukuba, 305-0801, Japan}

\begin{abstract}
We investigate the chiral anomaly in a Floquet system under a time-periodic electric field in (1+1) dimensions.  Using the van~Vleck high-frequency expansion, we analytically calculate the chiral current and the pseudo-scalar condensate for massless/massive fermions and how they are balanced with the topological charge.  In the high-frequency limit, we find that finite-mass effects are suppressed and the topological charge is dominated by the chirality production.  Our calculations show that the information about the chiral anomaly is stored not in the static Floquet Hamiltonian but in the periodic kick operator.  The computational steps are useful as the theoretical foundation for higher-dimensional generalization.
\end{abstract}


\maketitle

\tableofcontents

%%%%%%%%%%
\section{Introduction}

Chiral anomaly violates chiral symmetry in the classical Lagrangian through quantization~\cite{Fujikawa:1979ay,Fujikawa:1980eg}.
This anomalous violation of symmetry has received intensive attention since its discovery in a triangle diagram of $\pi^0\to 2\gamma$ decay~\cite{Adler:1969gk,Bell:1969ts}.
Among various applications, the anomaly-induced transport phenomenon attracts recent interest.
This is particularly important, for the chiral anomaly in the quantum sector causes a macroscopic observable which is detectable at the classical level.
Typical examples of anomaly-induced transport include the quantum Hall effect
and its counterpart with the chiral anomaly is the chiral magnetic effect.
In chiral systems, the chiral magnetic effect gives rise to a finite electric current along the magnetic field.
A similar effect known as the chiral separation effect exhibits charge separation along the magnetic field.
We note that such induced current and charge separation are prohibited by parity and time-reversal symmetries in classical electromagnetism.
In nuclear physics, a possibility of anomaly-induced observables has been pursued in the context of relativistic heavy-ion collisions~\cite{Kharzeev:2007tn,Kharzeev:2007jp,Fukushima:2008xe}.
Non-central collisions of heavy ions create a gigantic but transient magnetic field, and also gluonic configurations with nonzero topological charge density spread over transverse space, leading to local chiral imbalance through conversion from the topological charge density.
It is a critical question whether the chiral magnetic effect is really realized or not in the experimental setup~\cite{Kharzeev:2013ffa,Zhao:2019hta}.
For this purpose, nuclear isobar experiments aiming to extract the genuine signal from backgrounds by changing the magnetic strength have been conducted, as reported in Ref.~\cite{STAR:2021mii}.
Even though we need to make better digestion of data from the heavy-ion experiments, it is certainly true that anomalous transport phenomena are ubiquitous in various physics systems beyond nuclear field.
Weyl semimetals are characterized by fermionic excitation with the relativistic (i.e., nearly massless) dispersion relation~\cite{RevModPhys.90.015001}.
Then, a finite chiral imbalance can be injected externally through electromagnetic fields with $\bE\cdot\bB\neq 0$.
A signature for the chiral anomaly is the negative magneto-resistance~\cite{Son:2012bg}, and several experiments using Weyl semimetals have confirmed that resistivity could decrease with the increasing magnetic field as expected from theory~\cite{Son:2012bg,Li:2014bha,Lv:2015pya,Huang:2015eia}. 

For the chiral magnetic effect, the key issue is how to accumulate nonzero chiral imbalance.
In a setup with electromagnetic configurations with $\bE\cdot\bB\neq 0$, the background field pumps a finite magnetic helicity and the helicity is converted into nonzero chirality after all.
This conversion process is microscopically understood from the particle production under parallel $\bE$ and $\bB$.
Hence, the problem of pair particle production,
which can be traced back to a celebrated work of the Schwinger mechanism~\cite{PhysRev.82.664},
has been investigated also in the context of the chiral anomaly;
see Refs.~\cite{Dunne:2004nc,Fedotov:2022ely} for reviews.
The formula for the Schwinger mechanism is of great use to examine real-time dynamics of the chiral anomaly~\cite{Fukushima:2010vw,Warringa:2012bq,Copinger:2018ftr,Copinger_2020}.
In particular, there were controversial disputes about the absence of the chiral magnetic effect in equilibrium~\cite{PhysRevLett.111.027201,Yamamoto:2015fxa},
but they have been resolved by theoretical treatments in and out of equilibrium such as
the Schwinger mechanism~\cite{Fukushima:2010vw,Warringa:2012bq,Copinger:2018ftr,Copinger_2020},
the chiral kinetic theory~\cite{Stephanov:2012ki,Son:2012zy},
the anomalous hydrodynamics~\cite{Son:2009tf},
and the brute-force method of solving the Dirac equation~\cite{Fukushima:2020ncb,Aoi:2021azo}.
In most cases, constant (i.e., spatially homogeneous and time-independent) electromagnetic fields are assumed for technical reasons,
and only little is known about realistic situations with inhomogeneous and time-varying fields.
If the time dependence is steep enough and the frequency exceeds the mass gap threshold, the pair production dynamically occurs~\cite{Taya:2014taa}.
Such a perturbative process is qualitatively distinct from the Schwinger mechanism or quantum tunneling.
In this work, specifically, we focus on the high-frequency limit of time-dependent external fields.

In this work, we consider fast-oscillating electric fields and employ the Floquet theory to discuss the chiral anomaly for Weyl fermions and also {\it massive} Dirac fermions.  Namely, we shall explicitly calculate the chiral current and the pseudo-scalar condensate, which gives an extra contribution to the chiral-anomaly relation for the massive Dirac case, and discuss how they make up the chiral anomaly relation.  We emphasize that it is crucially important for the chiral-anomaly physics to separately and explicitly calculate the two contributions, since the anomaly relation is an equation for the total sum of the two and does not know the amount of each contribution.  Besides, what does matter in actual observables and the chiral-anomaly-induced phenomena such as the chiral magnetic effect is the chiral current, which cannot be obtained directly from the anomaly relation without explicit field-theoretical calculations.  As far as we are aware, this is the first study of the use of the Floquet theory to the chiral anomaly and also the massive chiral anomaly under a periodic driving.

The Floquet theory is a promising formulation to attack nonequilibrium phenomena under theoretical control~\cite{PhysRev.138.B979,PhysRevA.7.2203,PhysRevA.68.013820,Goldman:2014xja,Bukov_2015,Eckardt_2015,MANANGA20161,Oka_2019}.
The Floquet theorem states that the time evolution operator in periodically-driven systems can be decomposed into a time-independent effective Hamiltonian, called the Floquet Hamiltonian, and a periodic unitary operator, called the kick operator.  
Using the Floquet Hamiltonian, we can solve a static problem to identify the ``vacuum'' even though the original Hamiltonian varies with time.
When the frequency of external fields is higher than other typical scales, we can make use of a sophisticated method to construct the Floquet Hamiltonian as a power series of the inverse frequency.
As we briefly review later, the inverse frequency expansion still has a ``gauge'' ambiguity.
In the present work, we choose the van~Vleck expansion in which the time average of the kick operator vanishes by construction.

The idea of Floquet engineering is to adjust the periodic perturbation and design the Floquet Hamiltonian.  Then, an immediate question we would like to raise is the following:
starting from the periodically-driven system with the chiral anomaly, we can map the physics problem to the static system according to the Floquet theory.
How can we retain the (3+1)-dimensional chiral anomaly in the mapped static system?
We may well say that this question of chiral anomaly preservation over systems in different dimensions is a Floquet analog of the \textit{anomaly inflow}.

Our explicit calculations will reveal an interesting structure, i.e., the chiral anomaly is indeed washed out from static dynamics and is recovered solely from the kick operators at the edges of the time evolution.
In this paper, we dedicate ourselves to a simple version of the chiral anomaly in (1+1) dimensions.
There are two reasons for this choice.
First, we can solve the (1+1)-dimensional problem analytically without approximation.  Precisely speaking, for massless Weyl fermions in (1+1) dimensions, all quantities of our interest in this paper are analytically solvable.  Therefore, as a platform to test our ideas and methods, (1+1)-dimensional systems are simple but nontrivial enough.
Interestingly, the common technique of the high-frequency expansion in the Floquet theory can be continued to infinite order, and the full analytical structure will be transparently manifested.
Second, we would point out the possibility to realize (1+1)-dimensional systems experimentally.  When a strong magnetic field is applied to (3+1)-dimensional systems, the transverse motion is so restricted that the system effectively responds in (1+1) dimensions.  For example, in heavy-ion collisions where gigantic magnetic fields are transiently created, matter at the early stage should be subject to (1+1)-dimensional dynamics.  Also, the direct setup of (1+1)-dimensional systems is feasible with e.g.\ quantum wires as well as synthetic gauge fields.

In this paper, we first make a brief review of the chiral anomaly and the Floquet theory.  Then, we examine massless Weyl fermions and proceed to massive Dirac fermions.  The calculations for massless Weyl fermions show a profound structure of the high-frequency expansion.
That is, we find that the chiral anomaly is saturated by the first few terms in the expansion and the residual terms cancel out in the sum.
Therefore, to extract the chiral anomaly, we can simply neglect the higher-order terms even when the frequency is not large at all.
For massive Dirac fermions, the mass term and the pseudo-scalar condensate explicitly violate chiral symmetry, but the chiral anomaly is not lost.
The chiral anomaly itself is again captured by the first few terms of the chirality production as in the case of massless fermions.
Furthermore, the mass effect stems from the higher-order terms that do not cancel out.

In this work, we shall propose an interpretation of the chiral anomaly relation as the equation between the gauge and the matter sectors, where the latter is a sum of the chirality production and the pseudo-scalar condensate.  We discover that these two contributions behave very differently for electric fields with slow and rapid changes in time by explicitly calculating the two contributions separately using the Floquet theory for the first time.  Under slowly changing electric fields, the conventional Schwinger mechanism gives a clear explanation for the chiral anomaly due to quantum tunneling.
The chirality production through tunneling is exponentially suppressed, so the pseudo-scalar condensate dominates.  Meanwhile, in the high-frequency regime with rapid oscillations, the chirality production dominates.  There, the decay channel from energetic photons into fermions is open, and the chirality production is no longer exponentially suppressed.
For the chiral anomaly hunt in general, the mass effect always brings about the most subtle obstacle in interpretation.  Even for the Weyl semimetals a nonzero residual mass is unavoidable.
The present calculations would provide us with a clue to cope with the mass effect and probe the chiral anomaly in ideally designed environments.
\vspace{1em}

{\it Notations and conventions.---} We use the natural units, $c=\hbar=1$.  In this work we focus on (1+1)-dimensional spacetime:
$x^\mu = (x^0,x^1) \eqqcolon  (t,x)$, with a metric $g_{\mu\nu} \coloneqq {\rm diag}(+1,-1)$.  We occasionally use light-cone variables, $v^\pm$, which are defined as 
\begin{align}
    v^\pm \coloneqq \frac{v^0 \pm v^1}{\sqrt{2}} \eqqcolon  v_{\mp}\,.
\end{align}
We also define the Fourier transformation, $f(t,x) \to \tilde{f}_l(x)$, as 
\begin{align}
    \tilde{f}_l(x) \coloneqq \int^T_0 \frac{{\rm d}t}{T} f(t,x)\, {\rm e}^{-\frac{2\pi {\rm i}l}{T}t}\,,
\end{align}
where $T$ is the period (or the inverse frequency) of the Hamiltonian such that $H(t,x) = H(t+T,x)$.

%%%%%%%%%%
\section{Basics of the chiral anomaly and the Floquet theory}

To make the paper self-contained, we start with a brief review of the chiral anomaly and the Floquet theory based on the van~Vleck high-frequency expansion (see Refs.~\cite{Bukov_2015, MANANGA20161, Oka_2019} for more details).

%%%%%
\subsection{Chiral anomaly}
\label{subsec:review_anomaly}

Let us consider the massless Weyl fermions in (1+1) dimensions.  The essence of the chiral anomaly lies in the spectral flow of the Dirac sea in the presence of electric fields~\cite{Nielsen:1983rb}.  To explain this, we remind that in the massless (1+1)-dimensional case only two types of fermions exist, i.e., they move to either the right or the left with the linear dispersion, $E(p)= \pm p$.  The right- and left-moving fermions have right- and left-handed chirality, respectively.  In the absence of chemical potential, the Dirac sea is filled with negative-energy fermions up to $E=0$.  An external electric field accelerates the fermions to obtain momentum $\Delta p = \int^t {\rm d}t' eE(t')$.  Thus, fermions occupy the band spectra up to the energy level
$E \leq + \Delta p $ (right-handed)
and $E \leq -\Delta p$ (left-handed).
This means that the number of right-handed (and left-handed) fermions is increased (and decreased, respectively) by the amount of $\Delta N = \frac{\Delta p}{2\pi} = \frac{1}{2\pi} \int^t {\rm d}t' eE(t')$, that is nothing but the topological charge.
The differential representation of chirality change reads
$\partial_t N_{\rm R/L} = \pm eE/2\pi$.
In the covariant form, the chirality non-conservation is expressed as 
\begin{align}
	\partial_\mu J_{\rm R}^\mu 
	 = -\partial_\mu J_{\rm L}^\mu 
	 = -\frac{\varepsilon^{\mu\nu}F_{\mu\nu}}{4\pi} 
	 = \frac{eE}{2\pi} \,.
	 \label{eq:conv_anomaly11}
\end{align}
This is the chiral anomaly relation for Weyl fermions in (1+1) dimensions~\cite{Adler:1969gk,Bell:1969ts}.  

For massive Dirac fermions, the anomaly relation~\eqref{eq:conv_anomaly11} is modified by the finite mass contribution.  A nonvanishing Dirac mass makes right- and left-handed fermions coupled and explicitly breaks chiral symmetry, giving rise to an extra term in the anomaly relation as follows:
\begin{align}
	\partial_\mu J_{5}^\mu 
	 = \frac{eE}{\pi} + 2m P , \label{eq:conv_anomaly31}
\end{align}
where
$J_5^\mu \coloneqq J_{\rm R}^\mu - J_{\rm L}^\mu$
and
$P \coloneqq \braket{ \hat{\bar{\psi}}{\rm i}\gamma_5\hat{\psi} }$
with $\hat{\psi}$ the Dirac field.
We call $J_5^\mu$ the chiral current and $P$ the pseudo-scalar condensate.
Although Eq.~\eqref{eq:conv_anomaly31} is a conventional form,
we can reorganize terms so that the \textit{sum} of the divergence of the chiral current (which we call the chirality production, hereafter) and the pseudo-scalar condensate amounts to the topological charge density, $eE/\pi$.
This is a natural reorganization for the interpretation of the anomaly relation as the helicity conversion process from the gauge sector to the matter sector.
In other words, the anomaly relation is a \textit{sum rule} to quantify how the topological charge is saturated.
Thus, the sum rule does not care about the anatomy of each contribution in the matter sector.  To estimate each contribution to the sum rule, we must proceed to explicit computations.

In the slow limit case, where the spacetime variation of electric field is negligibly small, it is well understood how Eq.~\eqref{eq:conv_anomaly31} is saturated.  
The chirality production for massive Dirac fermions is then driven mainly by the Schwinger mechanism~\cite{Copinger_2020, Fukushima:2010vw, Tanji_2010, Warringa:2012bq, Taya:2020bcd, Aoi:2021azo, Ambjorn:1983hp}; that is, pairs of particles and anti-particles are produced via quantum tunneling.  The pair production number for slowly varying electric fields is estimated by Schwinger's classic calculation~\cite{Schwinger:1951nm} (in the so-called locally-constant-field approximation \cite{Bulanov:2004de, Dunne:2005sx}) as
\begin{align}
	N_{\rm pair}(t,x) \simeq \frac{1}{2\pi} \int^t {\rm d}t' eE(t',x)\; {\rm e}^{-\pi m^2/|eE(t,x)|} \,.  \label{eq::5}
\end{align}
After the pair production, the particles and the anti-particles are accelerated along the parallel and the anti-parallel directions with respect to the electric field, respectively, i.e., the spectral flow occurs.  Accordingly, the chirality is produced by the unit of $+2$ whenever a pair pops up from the vacuum.  Therefore,  
we expect
$J_5^0\simeq 2N_{\rm pair}$ in the slow limit.
This identification can be formulated in the covariant form, which is expressed as
\begin{align}
    \partial_\mu J_5^\mu \simeq \frac{eE(t,x)}{\pi}\; {\rm e}^{-\pi m^2/|eE(t,x)|} \,.  \label{eq::7}
\end{align}
We can also determine the pseudo-scalar condensate $P$ in the slow limit as
\begin{align}
	P \simeq -\frac{eE(t,x)}{\pi} \,\frac{ 1 - {\rm e}^{-\pi m^2/|eE(t,x)|}}{2m} \,. \label{eq:uu-88}
\end{align}
Therefore, the chiral anomaly in the slow-frequency limit can be saturated in the following way~\cite{Copinger:2018ftr}:
\begin{align}
    \underbrace{ \frac{eE}{\pi}}_{{\rm topological\ charge\ density}} \xrightarrow{{\rm slow\ limit}}  \underbrace{ \frac{eE}{\pi} {\rm e}^{-\pi m^2/|eE|}  }_{{\rm chirality\ production\ }\partial_\mu J^\mu_5} + \underbrace{ \frac{eE}{\pi} \left( 1 - {\rm e}^{-\pi m^2/|eE|} \right) }_{{\rm pseudo\mathchar`-scalar\ condensate\ }-2mP} \,. \label{eq:::9}
\end{align}

The massive chiral anomaly relation~\eqref{eq:conv_anomaly31} generally holds, but the splitting relation as discussed above depends on the detailed shapes of applied electric fields.
In fact, the estimate~\eqref{eq::7} is reasonable only in the slow limit.  When the electric field is rapidly varying in the high-frequency limit, the pair production is no longer driven by the Schwinger mechanism, but the production mechanism is similar to the photoelectric effect~\cite{Taya:2020dco, Brezin:1970xf, Dunne:2005sx, Keldysh:1965ojf,Popov:1971ff,Taya:2014taa}.  Moreover, when the frequency is high enough to be comparable to the typical lifetime of virtual particles, $\Delta t \sim 1/m$, real and virtual particles may not be clearly distinct.  Therefore, the physics picture of the spectral flow of real particles based on the Schwinger mechanism does not make sense in the high-frequency regime and additional contributions from virtual particles (e.g., the Zitterbewegung effect) need to be included.
After explicit calculations in the subsequent sections, we will come back to the question of how the chiral anomaly is saturated [see Eq.~\eqref{o56} for our final result].

\subsection{Floquet theory and van~Vleck expansion} \label{sec2.2}

We are interested in a periodic system such that
\begin{align}
	H(t)=H(t+T)\,.
\end{align}
The Floquet theorem~\cite{ASENS_1883_2_12__47_0} states that there exists a time-independent operator $H_{\rm F}$ called the Floquet Hamiltonian and a periodic operator $K(t)=K(t+T)$ called the kick operator, with which the unitary time-translation operator $U(t,t')$ (which we define as evolving the system from $t$ to $t'$) is expressed as
\begin{align}
	U(t,t') = {\rm e}^{-{\rm i}K(t')}\, {\rm e}^{-{\rm i}(t'-t)H_{\rm F}}\, {\rm e}^{+{\rm i}K(t)} \,.
\end{align}
We note that $H_{\rm F}^\dagger = H_{\rm F}$ and $K^\dagger = K$.
We can make use of perturbative expansion to solve $H_{\rm F}$ and $K$ from $-{\rm i}\partial_t U(t,t') =  U(t,t')H(t)$ and $1 = U(t,t)$ iteratively.  In the common assumption, the small expansion parameter is identified as the sufficiently small inverse frequency. 
 Namely, we add a book-keeping parameter $\varepsilon$ only in the explicit time derivative $\partial_t$ to count the order of the expansion, i.e.,
\begin{align}
	U(t,t') H(t)
		&= -{\rm i} \varepsilon^{-1} {\rm e}^{-{\rm i}K(t')} {\rm e}^{-{\rm i}(t'-t)H_{\rm F}} \frac{ \partial\, {\rm e}^{+{\rm i}K(t)} }{\partial t}  + {\rm e}^{-{\rm i}K(t')} H_{\rm F} {\rm e}^{-{\rm i}(t'-t)H_{\rm F}} {\rm e}^{+{\rm i}K(t)} \,.
\end{align}
Rearranging this expression, we find,
\begin{align}
	H_{\rm F} 
		= {\rm e}^{+{\rm i}K(t)}    H(t){\rm e}^{-{\rm i}K(t)} + {\rm i} \varepsilon^{-1} \frac{ \partial\, {\rm e}^{+{\rm i}K(t)} }{\partial t}  {\rm e}^{-{\rm i}K(t)} \,.  \label{xx172}
\end{align}
We aim to perform the power series expansion in the following form:
\begin{align}
	H_{\rm F}	\eqqcolon  \sum_{n=0}^\infty \varepsilon^n H_{\rm F}^{(n)}\,,\qquad
	K(t)		\eqqcolon  \sum_{n=0}^\infty \varepsilon^n K^{(n)}(t) \,. \label{xx9}
\end{align}
The expansion with respect to $\varepsilon$, which we refer to as the high-frequency expansion, can be conveniently regarded as the expansion using the period $T$; the time derivative $\partial_t$ picks up the frequency in Fourier space that is inversely proportional to $T$.  
The high-frequency expansion is thus expected to be reasonable if $1/T$ is sufficiently large compared to the other energy scales of the system.

To evaluate the series coefficients $H_{\rm F}^{(n)}$ and $K^{(n)}$, we use the following identities:
\begin{subequations}\label{eq-10}
\begin{align}
	{\rm e}^{+{\rm i}K(t)}H(t){\rm e}^{-{\rm i}K(t)}
		&= H + {\rm i}[K,H] - \frac{1}{2}[K,[K,H]] - \frac{{\rm i}}{6} [K,[K,[K,H]]] + {\mathcal O}(K^4) \,, \\
	\frac{ \partial\, {\rm e}^{+{\rm i}K(t)} }{\partial t}  {\rm e}^{-{\rm i}K(t)} 
		&= {\rm i}\dot{K} - \frac{1}{2}[K,\dot{K}] - \frac{{\rm i}}{6}[K,[K,\dot{K}]] + \frac{1}{24}[K,[K,[K,\dot{K}]]] + {\mathcal O}(K^5) \,.
\end{align}
\end{subequations}
Then, the lowest-order term of ${\mathcal O}(\varepsilon^{-1})$ in Eq.~(\ref{xx172}) gives
\begin{align}
	0 = \dot{K}^{(0)} \qquad \Rightarrow \qquad
	K^{(0)} = {\rm const.}
\end{align}
This constant is arbitrary, in principle, and can be understood as ``gauge ambiguity'' related to the arbitrariness of the origin of time.  To fix the constant, the van~Vleck expansion requires,
\begin{align}
	\int^T_0 {\rm d}t\,K(t) = 0 \,, \label{vv178}
\end{align}
which yields $K^{(0)} = 0$.
Note that the physical results are independent of the Floquet gauge fixing, and different choices like the Floquet-Magnus gauge fixing condition with $K(0)=0$ lead to the Floquet Hamiltonian $H_{\rm F}$ and the kick operator $K$,
which are related to the van~Vleck results through an appropriate Floquet gauge transformation~\cite{Bukov_2015}.
Thanks to $K^{(0)}=0$, we can identify ${\mathcal O}(K^{(n)}) = {\mathcal O}(\varepsilon^n)$ and can expand Eq.~(\ref{xx172}) to obtain,
\begin{subequations} \label{a182}
\begin{align}
	H_{\rm F}^{(0)} 
		&= H - \dot{K}^{(1)} \,,  \label{a182a} \\
	H_{\rm F}^{(1)} 
		&= +{\rm i}[K^{(1)},H] - \dot{K}^{(2)} -\frac{{\rm i}}{2}[K^{(1)},\dot{K}^{(1)}]  \,, \label{a182b}\\
	H_{\rm F}^{(2)} 
		&= + {\rm i}[K^{(2)},H] -\frac{1}{2}[K^{(1)},[K^{(1)},H]] - \dot{K}^{(3)} -\frac{{\rm i}}{2}[K^{(1)}, \dot{K}^{(2)}] \nonumber\\
		&\quad -\frac{{\rm i}}{2}[K^{(2)},\dot{K}^{(1)}]  +\frac{1}{6}[K^{(1)},[K^{(1)},\dot{K}^{(1)}]]  \label{a182c} 
\end{align}
\end{subequations}
up to ${\mathcal O}(\varepsilon^3)$.  Equation~(\ref{a182}) can be solved uniquely under the Floquet gauge fixing condition~(\ref{vv178}), and the first few terms of $H^{(n)}_{\rm F}$ and $K^{(n)}$ in terms of the given Hamiltonian $H$ read~\cite{Bukov_2015, Goldman:2014xja, Mizuta_2019}
\begin{subequations} \label{oo16}
\begin{align}
	H^{(0)}_{\rm F} &= \tilde{H}_0 \,, \\
	H_{\rm F}^{(1)} &= \frac{{\rm i}}{2} \sum_{l\neq0} \frac{T}{2\pi {\rm i} l} \left[ \tilde{H}_l , \tilde{H}_{-l} \right] \,, \\
	H_{\rm F}^{(2)} &=  \frac{1}{2}  \sum_{l\neq 0}   \left(\frac{T}{2\pi {\rm i} l}\right)^2  \left[  \tilde{H}_{-l} , \left[ \tilde{H}_l , \tilde{H}_0 \right]  \right] - \frac{1}{3} \sum
_{l\neq 0} \sum_{l'\neq 0,-l}   \frac{T}{2\pi {\rm i} l'} \frac{T}{2\pi {\rm i} l}   \left[ \tilde{H}_{l} ,  \left[ \tilde{H}_{l'} , \tilde{H}_{-l-l'} \right]\right] 	\,,
\end{align}
\end{subequations}
and
\begin{subequations} \label{oo17}
\begin{align}
	K^{(0)}(t) 
		&= 0 \,, \\
	K^{(1)}(t) 
		&= \sum_{l\neq 0} \frac{T}{2\pi {\rm i} l}{\rm e}^{{\rm i}\frac{2\pi l}{T}t} \tilde{H}_{l} \,, \\
	K^{(2)}(t) 
		&= +\frac{{\rm i}}{2} \sum_{l\neq 0} \sum_{l'\neq 0,l} \frac{T}{2\pi {\rm i} l}\frac{T}{2\pi {\rm i} l'}{\rm e}^{{\rm i}\frac{2\pi l}{T}t} \left[ \tilde{H}_{l'} , \tilde{H}_{l-l'} \right] + {\rm i} \sum_{l\neq 0} \left(\frac{T}{2\pi {\rm i} l}\right)^2{\rm e}^{{\rm i}\frac{2\pi l}{T}t}   \left[ \tilde{H}_l , \tilde{H}_0 \right] \,.   
\end{align}
\end{subequations}
We can continue the expansion to higher-order terms with $n\geq 3$ systematically.

%%%%%%%%%%
\section{Weyl fermions} \label{sec3}

We first consider a simple case of the Weyl fermions in (1+1) dimensions to make it clear how the van~Vleck high-frequency expansion in the Floquet theory can correctly reproduce the chiral anomaly and explain for the first time how the chiral anomaly relation can be understood in terms of the Floquet theory.
We then proceed to discussions for a more nontrivial case of the massive Dirac fermions in the next section.
Using the Weyl fermions in (1+1) dimensions is advantageous for theoretical consideration since the Weyl equation is analytically solvable for any background fields.
We can therefore write down the explicit forms of the Floquet Hamiltonian $H_{\rm F}$ and the kick operator $K$ to arbitrary orders.
We show that the chiral anomaly in the van~Vleck high-frequency expansion has an intriguing structure;
only low-order terms up to $n=2$ encompass the chiral anomaly and higher-order terms with $n>2$ eliminate redundant terms in the previous order.
We also clarify the convergence condition for the van~Vleck expansion, and discuss how we can extend the radius of convergence via resummation or considering a ``Hamiltonian'' in the spatial direction.  

%%%%%
\subsection{Weyl equation and Hamiltonian} \label{sec3.1}

The Weyl equation in (1+1) dimensions reads,
\begin{align}
	{\rm i} \partial_t\, \chi_{\rm R/L}(t,x) = H_{\rm R/L}(t,x)\, \chi_{\rm R/L}(t,x) \label{v1}
\end{align}
with the Weyl Hamiltonian expressed in terms of the light-cone variables,
\begin{align}
	H_{\rm R/L}	\coloneqq \mp{\rm i}\partial_x + eA^0(t,x) \mp eA^1(t,x)  
		= \mp{\rm i}\partial_x + \sqrt{2} eA^\mp(t,x)\,.
    \label{eq:Weyl-H}
\end{align}
The upper and the lower signs refer to the right- and the left-handed [or the right- and the left-moving in (1+1) dimensions] fermions, i.e., $\chi_{\rm R}$ and $\chi_{\rm L}$, respectively.  We will fix the gauge later and the gauge background $A^\mu$ is not in any particular gauge yet for the moment.
In our problem setup the electric field is periodic in time and thus $A^\mu$ satisfies,
\begin{align}
	A^\mu(t+T,x) = A^\mu(t)\,,
\end{align}
so as the Hamiltonian; $H(t+T,x)=H(t,x)$.
We may then express $H_{\rm R/L}$ in Eq.~(\ref{eq:Weyl-H}) as a Fourier series as
\begin{align}
	H_{\rm R/L}
    = \tilde{H}_0 + \sum_{l\neq 0} {\rm e}^{+\frac{2\pi {\rm i}l}{T}t} \tilde{H}_{l\neq 0}
    = \mp{\rm i}\partial_x + \sqrt{2}e\tilde{A}^\mp_0 \;+\; \sum_{l\neq 0} {\rm e}^{+\frac{2\pi{\rm i}l}{T}t} \sqrt{2}e\tilde{A}^\mp_l \,.
\end{align}
Here, the first two terms in the right-hand side constitute the zero mode $\tilde{H}_0$.
We also assume that the period $T$ is sufficiently small compared to the typical length scale of the electric field $L$, i.e.,
\begin{align}
	T/L \ll 1 \,,
\end{align}
so that the van~Vleck high-frequency expansion can converge as we shall demonstrate below.

\subsection{Van~Vleck expansion to arbitrary orders} \label{sec3.2}

We can compute the Floquet Hamiltonian $H_{\rm F}$ and the kick operator $K$ to arbitrary orders for the Weyl equation in (1+1) dimensions.  As reviewed in Sec.~\ref{sec2.2}, the Floquet Hamiltonian $H_{\rm F}$ and the kick operator $K$ are obtained by perturbatively solving Eq.~(\ref{xx172}).  For this, we notice that only the zero mode of the Hamiltonian, $\tilde{H}_0 = \mp {\rm i}\partial_x+\sqrt{2}e\tilde{A}^\mp_0$, can have nontrivial commutation relations, which results in only the c-numbers (no derivative terms).
Therefore, the kick operator $K$ is a c-number as well and commutes with anything except for $\tilde{H}_0$.
We can simplify Eq.~(\ref{eq-10}) as
\begin{align}
	{\rm e}^{+{\rm i}K}H{\rm e}^{-{\rm i}K}
		= H + {\rm i}[K,\tilde{H}_0] \,, \qquad
	\frac{ \partial\, {\rm e}^{+{\rm i}K} }{\partial t} {\rm e}^{-{\rm i}K} 
		&= {\rm i}\dot{K} \,,
\end{align}
which in turn simplifies Eq.~(\ref{xx172}) as
\begin{align}
	H_{\rm F} = H + {\rm i}[K,\tilde{H}_0] - \varepsilon^{-1}\dot{K} \,.    \label{aa24}
\end{align}
Applying $\int^T_0\frac{{\rm d}t}{T}$ and imposing the van~Vleck gauge fixing condition~(\ref{vv178}), we arrive at
\begin{align}
	H_{\rm F} = \tilde{H}_0 \,,   \label{oo24}
\end{align}
i.e., there are no higher order corrections to the Floquet Hamiltonian $H_{\rm F}$ within the van~Vleck expansion scheme.  We then understand that $K^{(n)}$ satisfies a recursive equation as follows:
\begin{align}
	\dot{K}^{(n+1)} = {\rm i}[K^{(n)},\tilde{H}_0] = \mp \partial_x K^{(n)}  \label{pp26}
\end{align}
with an initial condition given by
\begin{align}
	\dot{K}^{(1)} = H - \tilde{H}_0 = \sqrt{2} \left( eA^{\mp} - e\tilde{A}^{\mp}_0 \right)\,.
\end{align}
The solution to this recursive equation is found to be
\begin{align}
	K^{(n)}(t,x) = (\mp \partial_x)^{n-1} \int^t {\rm d}t_{n} \int^{t_n} {\rm d}t_{n-1} \cdots \int^{t_2} {\rm d}t_{1} \sqrt{2} \left( eA^{\mp}(t_1,x) - e\tilde{A}^{\mp}_0(x) \right) \,. \label{pp27}
\end{align}
Throughout this work we write indefinite integrals dropping undetermined integration constants for notational simplicity.
For example, $\int^t {\rm d}t' {\rm e}^{-\frac{2\pi{\rm i}}{T}t'} \coloneqq \frac{T}{-2\pi{\rm i}} {\rm e}^{-\frac{2\pi{\rm i}}{T}t'}$ in our convention.
Note that the integration constant must be vanishing in Eq.~(\ref{pp27}) to satisfy the van~Vleck gauge fixing condition~(\ref{vv178}).  Summing up with respect to $n$, we find,
\begin{align}
	K(t,x) &= \sum_{n=1}^\infty (\mp \partial_x)^{n-1} \int^t {\rm d}t_{n} \int^{t_n} {\rm d}t_{n-1} \cdots \int^{t_2} {\rm d}t_{1} \sqrt{2}  \left( eA^{\mp}(t_1,x) - e\tilde{A}^{\mp}_0(x) \right) \nonumber\\
		&= \sum_{n=1}^\infty (\mp \partial_x)^{n-1} \int^t {\rm d}t' \frac{(t-t')^{n-1}}{(n-1)!} \sqrt{2}  \left( eA^{\mp}(t',x) - e\tilde{A}^{\mp}_0(x) \right) \nonumber\\
		&= \int^t {\rm d}t' \sqrt{2}  \left( eA^{\mp}(t',x\mp (t-t')) - e\tilde{A}^{\mp}_0(x \mp (t-t')) \right) \nonumber\\
		&= \int^{x^\pm} {\rm d}x^{\prime \pm}  \left[ eA^{\mp}\left( \frac{x^{\prime \pm} + x^{\mp}}{\sqrt{2}},  \pm \frac{x^{\prime \pm} - x^{\mp}}{\sqrt{2}} \right) - e\tilde{A}^{\mp}_0\left( \pm \frac{x^{\prime \pm} - x^{\mp}}{\sqrt{2}} \right) \right] \,, \label{ww30}
\end{align}
where we used an identity, $\int^t {\rm d}t_{n} \int^{t_n} {\rm d}t_{n-1} \cdots \int^{t_2} {\rm d}t_{1} f(t_1) = \int^t {\rm d}t' \frac{(t-t')^{n-1}}{(n-1)!} f(t')$ from the first to the second line, and changed the integration variable $t' \to x^{\prime \pm} \coloneqq \sqrt{2}t'-x^{\mp}$ to get the last line.
Also, we omitted the book-keeping parameter $\varepsilon$ for brevity.

We emphasize that $K$ receives nontrivial corrections with $A^\mu$, while $H_{\rm F}$ does not.
This already implies that nontrivial physics such as the chiral anomaly comes out from the kick operator.
Interestingly, the series expression in the first line has a finite radius of convergence, which means that the van~Vleck expansion is not necessarily convergent.  Nevertheless, once the resummation with respect to $n$ is performed, the final expression can be analytically continued beyond the original radius of convergence and is well-defined on the whole parameter space.
This is an example of how we can extend the finite radius of convergence via resummation in the van~Vleck high-frequency expansion.
Note that the Weyl equation in (1+1) dimensions is analytically solvable and this enables us to
obtain $K^{(n)}$'s to arbitrary orders and
take the summation exactly here.
In general, it is a difficult task to compute $K^{(n)}$'s for large $n$.  In such cases, we should resort to some approximations, e.g., P\'{a}de approximation, Borel-P\'{a}de approximation, etc.\ to carry out the resummation.
We also note that we can directly obtain the resummed expression~(\ref{ww30}) by directly solving the differential equation~(\ref{aa24}) instead of using the recursive relation~(\ref{pp26}).  Indeed, Eq.~(\ref{oo24}) simplifies Eq.~(\ref{aa24}) as $\partial_\pm K = eA^\mp - e\tilde{A}^\mp_0$, whose exact solution agrees with Eq.~(\ref{ww30}).

We continue discussions about the convergence condition in the van~Vleck expansion for the Weyl-fermion case (\ref{ww30}).  For simplicity, we take the axial gauge $A^1=0$ for the moment\footnote{We may estimate the magnitude as $A^1 = {\mathcal O}(eET)$, which is different by $T/L$ compared to $A^0$.  Accordingly, in a different gauge choice in which $A^1$ is nonvanishing, the power counting of $K^{(n)}$ becomes cumbersome, though our final conclusion is not modified.  }.  We may estimate the typical magnitude as $A^0 = {\mathcal O}(eEL)$, where $L$ is the typical length scale for the electric field variation, and accordingly $K^{(n)}$ as $K^{(n)} = {\mathcal O}(A^0 T^n/L^{n-1}) = {\mathcal O}(eET^n/L^{n-2})$.  We may thus estimate the magnitude of the ratio between successive $K^{(n)}$'s as 
\begin{align}
	\left| \frac{K^{(n)}}{K^{(n-1)}} \right| = {\mathcal O}(T/L) \,.
\end{align}
According to d'Alembert's criterion, the van~Vleck expansion is convergent for
\begin{align}
	T/L < 1 \,. \label{gg31}
\end{align}
%This is consistent with our naive expectation~(\ref{pp9}).
Although the van~Vleck expansion fails to converge for $T/L > 1$, as mentioned above, a remedy for $T/L \geq 1$ is to complete the resummation for $T/L<1$ and perform the analytical continuation.
An alternative resolution is to consider the van~Vleck expansion in the $x$-direction instead of the $t$-direction.
Namely, we first rewrite the Weyl equation~(\ref{v1}) as
\begin{align}
	{\rm i}\partial_x \chi = \left[ \mp {\rm i}\partial_t \pm \sqrt{2}eA^\mp \right] \chi \eqqcolon  {\mathcal H} \chi\,.
\end{align}
We then apply the Floquet theory to the ``Hamiltonian'' ${\mathcal H}$ under the periodic condition in the $x$-direction, i.e., ${\mathcal H}(t,x)={\mathcal H}(t,x+L)$.  Repeating similar calculations, we can readily show that the corresponding Floquet Hamiltonian ${\mathcal H}_{\rm F}$ and the kick operator ${\mathcal K}$ are given by
\begin{subequations} \label{eq-31}
\begin{align}
	&{\mathcal H}_{\rm F} 
		= \bar{\mathcal H}_0 \,,\\
	&{\mathcal K} (t,x)
		= \pm \sum_{n=1}^{\infty} (\mp \partial_t)^{n-1} \int^x {\rm d}x_{n} \int^{x_n} {\rm d}x_{n-1} \cdots \int^{x_2} {\rm d}x_{1} \sqrt{2} \left( eA^{\mp}(t,x_1) - e\bar{A}^{\mp}_0(t) \right) \,,
\end{align}
\end{subequations}
where the Fourier transformation is applied in the spatial direction, $\bar{f}_l(t) \coloneqq \int^L_0 \frac{{\rm d}x}{L} f(t,x)\,{\rm e}^{-\frac{2\pi {\rm i}l}{L}x} $.
An essential change here compared to the van~Vleck expansion in the $t$-direction~(\ref{ww30}) is that the roles of $t$ and $x$ are interchanged.  The power counting is modified accordingly and the van~Vleck expansion in the $x$-direction~(\ref{eq-31}) leads to a convergent series for $T/L>1$.
In general, for systems having more than one coordinate variable, we may identify the appropriate linear combination of the variables with which the Hamiltonian varies the most rapidly.

Coming back to the expansion along the $t$-direction, we can construct the wavefunction $\chi$ using the full expressions for $H_{\rm F}$ in Eq.~(\ref{oo24}) and $K$ in Eq.~(\ref{ww30}).  To do this, it is convenient to introduce,
\begin{align}
	\phi (x) \coloneqq {\rm e}^{+{\rm i}K(0,x)} \chi (0,x) \,,  \label{ff22}
\end{align}
and choose $\phi$ to be an eigenfunction of the Floquet Hamiltonian~(\ref{oo24}) as
\begin{align}
	\lambda \phi(x) = H_{\rm F}(x)\, \phi(x) \,. \label{ff33}
\end{align}
Note that we may interpret Eq.~(\ref{ff22}) as an initial condition for the wavefunction at $t=0$.  The solution to Eq.~(\ref{ff33}) is
\begin{align}
	\phi_p(x) = \frac{{\rm e}^{\pm {\rm i}px}}{\sqrt{2\pi}} \exp\left[ \mp {\rm i}\int^x {\rm d}x' \sqrt{2} e\tilde{A}^\mp_0 (x') \right] \,,
\end{align}
which is labeled by kinetic momentum $p$.
The corresponding energy eigenvalue $\lambda$ is simply
$\lambda = p$.
Note that we normalized $\phi$ as $ \delta(p-p') = \int {\rm d}x\, \phi_p^\dagger\phi_{p'} = \int {\rm d}x\, \chi_p^\dagger\chi_{p'}$.  Now, by applying the time-translation operator $U$ onto $\chi_p(0,x)$, we obtain,
\begin{align}
	\chi_{p}(t,x) 
		&= U(0,t;x) \chi_{p}(0,x) \nonumber\\
		&= {\rm e}^{-{\rm i}K(t,x)} {\rm e}^{-{\rm i}H_{\rm F}(x)t} \phi_{p}(x) \nonumber\\
		&= \exp\left[ -{\rm i}\sum_{n=1}^\infty (\mp \partial_x)^{n-1} \int^t {\rm d}t_{n} \int^{t_n} {\rm d}t_{n-1} \cdots \int^{t_2} {\rm d}t_{1} \sqrt{2} \left( eA^{\mp}(t_1,x) - e\tilde{A}^{\mp}_0(x) \right) \right] {\rm e}^{-{\rm i}pt} \phi_{p}(x) \,.  \label{ll38}
\end{align}
We may explicitly carry out the $n$-summation in Eq.~(\ref{ll38}), as done in Eq.~(\ref{ww30}).  The result is,
\begin{align}
	\chi_{p}(t,x) 
		&=  \frac{{\rm e}^{-{\rm i}p \sqrt{2}x^\mp}}{\sqrt{2\pi}} \exp \left[ -{\rm i}\int^{x^\pm} {\rm d}x^{\prime \pm} eA^\mp  \right] \,,
\end{align}
which reproduces the exact solution to the Weyl equation~(\ref{v1}) in (1+1) dimensions, i.e., $0 = \left[ {\rm i}\partial_\pm - eA^\mp \right] \chi$.  

We can canonically quantize the field $\chi \to \hat{\chi}$ via the standard canonical quantization procedure.  We then expand the field operator $\hat{\chi}$ in terms of the wavefunctions~(\ref{ll38}) to define creation and annihilation operators as
\begin{align}
	\hat{\chi} 
		\eqqcolon  \int^{+\infty}_{-\infty} {\rm d}p\,  \chi_p \hat{a}_p 
		= \int^{+\infty}_{0} {\rm d}p \left[ \chi_p \hat{a}_p + \chi_{-p} \hat{b}_{p}^\dagger \right] \,, \label{rr41}
\end{align}
where $\hat{b}_{p}^\dagger \coloneqq \hat{a}_{-p}$ (with $p>0$ for Weyl fermions).  Since $\chi_{p}$ (and $\chi_{-p}$) has positive (and negative) energy eigenvalue $\lambda = p$ (and $\lambda=-p$, respectively), we may identify $\hat{a}_p$ and $\hat{b}_{p}^\dagger$ as annihilation operators for a Weyl fermion and a creation operator for an anti-Weyl fermion, respectively.  Accordingly, the vacuum state $\ket{0}$ is defined by
\begin{align}
	0 = \hat{a}_p \ket{0} = \hat{b}_p \ket{0}    \label{gg42}
\end{align}
for any positive $p$.
The commutation relations for the annihilation operators $\hat{a}_p$ and $\hat{b}_{p}$ read
\begin{align}
	\delta(p-p') = \{ \hat{a}_p, \hat{a}_{p'}^\dagger \} = \{ \hat{b}_p, \hat{b}_{p'}^\dagger \}\,,\qquad ({\rm others}) = 0 \,.   \label{gg43}
\end{align}
Note that the Floquet Hamiltonian $H_{\rm F}$ is independent of time, implying that a system under periodic driving may be regarded as a steady state.  Thus, we may naturally define a steady ``vacuum'' state, as in Eq.~(\ref{gg42}), via the creation/annihilation operators~(\ref{rr41}), though they are defined for the eigenstates of the Floquet Hamiltonian $H_{\rm F}$.

\subsection{Chiral current}

We compute the vacuum expectation value of the chiral current $J_5^\mu$ based on the all-order van~Vleck results obtained previously in Sec.~\ref{sec3.2}.
In general, expectation values suffer from ultraviolet  (UV) divergences, which should be regularized in a physically meaningful way.  As a regularization scheme, we adopt the point-splitting regularization scheme, which manifestly preserves the gauge invariance.  Namely, we define the chiral current $J_5^\mu$ as 
\begin{align}
    J_5^\mu \coloneqq J_{\rm R}^\mu - J_{\rm L}^\mu
    \label{ff44}
\end{align}
where $J_{{\rm R}}^\mu$ and $J_{\rm L}^\mu$ are contributions from right- and left-handed fermions, respectively, and are given by 
\begin{align}
    J_{{\rm R}}^\mu \coloneqq \begin{pmatrix} N_{{\rm R}} \\ N_{{\rm R}}   \end{pmatrix}\ \ {\rm and}\ \ 
    J_{{\rm L}}^\mu \coloneqq \begin{pmatrix} N_{{\rm L}} \\ -N_{{\rm L}}   \end{pmatrix}
\end{align}
with 
\begin{align}
	N_{{\rm R}/{\rm L}}
        &\coloneqq \braket{\hat{N}_{{\rm R}/{\rm L}}}(x^\mu) \nonumber\\
	  &\coloneqq \braket{\hat{\chi}^\dagger_{{\rm R}/{\rm L}} \hat{\chi}_{{\rm R}/{\rm L}}}(x^\mu) \nonumber\\
	  &\coloneqq \lim_{\epsilon^\mu \to 0} {\rm e}^{-{\rm i}\int^{x^\mu+\epsilon^\mu/2}_{x^\mu-\epsilon^\mu/2}{\rm d}x^\nu eA_\nu } \braket{0| \hat{\chi}_{{\rm R}/{\rm L}}^\dagger(x^\mu+\epsilon^\mu/2) \hat{\chi}_{{\rm R}/{\rm L}}(x^\mu-\epsilon^\mu/2) |0} \,. \label{ff45}
\end{align} 
Unless needed, we do not specify the subscripts ${\rm R}$ and ${\rm L}$ in what follows.  In the point-splitting regularization~(\ref{ff45}), the Wilson-line factor ${\rm e}^{-{\rm i}\int^{x^\mu+\epsilon^\mu/2}_{x^\mu-\epsilon^\mu/2}{\rm d}x^\nu eA_\nu }$ is inserted so as to guarantee the gauge invariance.  It is important to note that the $\epsilon^\mu\to 0$ limit should be taken symmetrically with respect to the Lorentz indices, so that expectation values can be covariant under Lorentz transformation~\cite{Peskin:1995ev}; this requires,
\begin{align}
	\lim_{\epsilon^\mu \to 0} \frac{\epsilon^\mu\epsilon^\nu}{\epsilon^2} = \frac{1}{2} g^{\mu\nu} \,,
\end{align}
for example.
Now, using the mode expansion~(\ref{rr41}), we may evaluate the expectation value~(\ref{ff45}) explicitly as
\begin{align}
	N
		&= \lim_{\epsilon^\mu \to 0} \int^{+\infty}_{0} \frac{{\rm d}p}{2\pi} {\rm e}^{-{\rm i}p \sqrt{2} \epsilon^{\mp}} \left[ 1 -{\rm i} \epsilon^\mu eA_\mu \pm {\rm i} \epsilon^1 \sqrt{2} e\tilde{A}^\mp_0 + {\rm i}\epsilon^\mu \partial_\mu K + {\mathcal O}(|\epsilon|^2)\right] \nonumber\\
		&= \frac{1}{2\pi} \lim_{\epsilon^\mu \to 0} \frac{-{\rm i}}{\sqrt{2}}\frac{1}{ \epsilon^{\mp}} \left[ 1 -{\rm i} \epsilon^\mu eA_\mu \pm {\rm i} \epsilon^1 \sqrt{2} e\tilde{A}^\mp_0 + {\rm i}\epsilon^\mu \partial_\mu K + {\mathcal O}(|\epsilon|^2)\right] \nonumber\\
		&= \frac{1}{2\pi\sqrt{2}}\left[ ({\rm const.\ div.}) - eA^\pm - e\tilde{A}^\mp_0 + \partial_\mp K \right] \,,
\end{align}
where and hereafter the upper and lower signs and indices refer to the right- and the left-handed fermions, respectively.  We remark that the time derivative $\partial_t$ appears from the point-splitting regularization procedure.  This means that, if we compute $N=\langle\hat{N}\rangle$ using the van~Vleck expansion truncated at a certain order $n=n_{\rm max}$, the computed $J$ becomes an expansion truncated at $n=n_{\rm max}-1$ because of $\partial_t = {\mathcal O}(\varepsilon^{-1})$.  Since physical currents should be zero in the absence of electric fields, it is physically sensible to subtract the constant divergence.  After the subtraction $N \to \ [N]$ (we enclose the expectation value by square brackets to denote the subtracted quantity) and substituting Eq.~(\ref{ww30}), we obtain,
\begin{align}
	[N] 
		= \frac{1}{2\pi\sqrt{2}}\left[  - eA^\pm - eA^\mp + 2\sum_{n=0}^\infty (\mp \partial_x)^{n} \int^t {\rm d}t_{n} \cdots \int^{t_2} {\rm d}t_{1} \left( eA^{\mp}(t_1,x) - e\tilde{A}^{\mp}_0(x) \right) \right] \,, \label{rr48}
\end{align}
where in the summation with respect to $n$ above, $\int^t {\rm d}t_{n} \cdots \int^{t_2} {\rm d}t_{1} \left( eA^{\mp}(t_1,x) - e\tilde{A}^{\mp}_0(x) \right)$ for $n=0$ should be understood as $eA^{\mp} - e\tilde{A}^\mp_0$.  Note that, similarly to Eq.~(\ref{ww30}), it is possible to carry out the $n$-summation to find,
\begin{align}
	[N] 
		= \frac{1}{2\pi\sqrt{2}}\left[  - eA^\pm + \int^{x^\pm}{\rm d}x^\pm \partial_\mp eA^\mp \right]  \,.   \label{rr49}
\end{align}
The resummed expression (\ref{rr49}) can be analytically continued to any values of $T/L$, while the series expression~(\ref{rr48}) converges only for $T/L < 1$.

\subsection{Chiral anomaly}

We proceed to the chiral anomaly relation based on the van~Vleck high-frequency expansion.  Taking the total derivative of the current $J^\mu$, we immediately recover,
\begin{align}
	\partial_\mu [J^\mu] 
		&= (\partial_t \pm \partial_x ) [N] \nonumber\\
		&= \frac{1}{2\pi}\partial_\pm \left[  - eA^\pm - eA^{\mp} + 2\sum_{n=0}^\infty (\mp \partial_x)^{n} \int^t {\rm d}t_{n} \cdots \int^{t_2} {\rm d}t_{1} \left( eA^{\mp}(t_1,x) - e\tilde{A}^{\mp}_0(x) \right) \right]  \nonumber\\
		&= \pm \frac{eE}{2\pi} \,, \label{yy50}
\end{align}
where $E \coloneqq F_{01} = F_{-+}$ is the electric field, and hence $\partial_\mu [J^\mu_5] = eE/\pi$.  This is nothing but the chiral anomaly relation for Weyl fermions in (1+1) dimensions.  

The interesting structure in Eq.~(\ref{yy50}) is that
the higher order terms with $n>2$ cancel with each other.
Thus, it is sufficient to truncate the van~Vleck expansion at $n=2$ to capture the full information about the chiral anomaly relation for Weyl fermions in (1+1) dimensions,
which motivates general speculation that the chiral anomaly could be dominantly encoded in the low-order terms in the high-frequency expansion in the Floquet theory.  To see the cancellation explicitly in the present case, let us truncate the series expression for the kick operator $K$ in Eq.~(\ref{ww30}) at $n=n_{\rm max} \geq 2$, which yields,
\begin{align}
	K = \sum_{n=1}^{n_{\rm max}} \varepsilon^n (\mp \partial_x)^{n-1} \int^t {\rm d}t_{n} \int^{t_n} {\rm d}t_{n-1} \cdots \int^{t_2} {\rm d}t_{1} \sqrt{2}  \left( eA^{\mp}(t_1,x) - e\tilde{A}^{\mp}_0(x) \right) + {\mathcal O}(\varepsilon^{n_{\rm max}+1}) \,,  
\end{align}
where we took the book-keeping parameter $\varepsilon$ back.  Correspondingly, the truncated number density $[N]$ reads, 
\begin{align}
	[N] 
		= \frac{1}{2\pi\sqrt{2}} \left[  - eA^\pm - eA^{\mp} + 2\sum_{n=0}^{N-1} \varepsilon^n (\mp \partial_x)^{n} \int^t {\rm d}t_{n} \cdots \int^{t_2} {\rm d}t_{1} \left( eA^{\mp}(t_1,x) - e\tilde{A}^{\mp}_0(x) \right) \right] + {\mathcal O}(\varepsilon^{n_{\rm max}}) \,.
\end{align}
As stressed before, the truncation lowers the order as ${\mathcal O}(\varepsilon^{n_{\rm max}+1}) \to {\mathcal O}(\varepsilon^{n_{\rm max}})$ because of $\partial_t = {\mathcal O}(\varepsilon^{-1})$ involved in the point-splitting regularization.  Then, we can evaluate the total derivative as
\begin{align}
	\partial_\mu [J^\mu] 
		&= ( \varepsilon^{-1}\partial_t \pm \partial_x ) \frac{1}{2\pi\sqrt{2}} \nonumber\\
			&\quad \times \left[  - eA^\pm - eA^{\mp} + 2\sum_{n=0}^{n_{\rm max}-1} \varepsilon^n (\mp \partial_x)^{n} \int^t {\rm d}t_{n} \cdots \int^{t_2} {\rm d}t_{1} \left( eA^{\mp}(t_1,x) - e\tilde{A}^{\mp}_0(x) \right) \right] + {\mathcal O}(\varepsilon^{n_{\rm max}-1}) \nonumber\\
		&= \pm \frac{E}{2\pi} + {\mathcal O}(\varepsilon^{n_{\rm max}-1}) \,.
\end{align}
It is easy to observe the general cancellation pattern that the term acted by $\partial_x$ is exactly eliminated by the next order term acted by $\varepsilon^{-1}\partial_t$.
The nontrivial feature we would like to emphasize is that both $K$ and $[J^\mu]$ receive higher-order corrections,
while the cancellation occurs only after taking $\partial_\mu$ on $[J^\mu]$,
which reflects the fact that the anomaly relation is exact.

\section{Massive Dirac fermions} \label{sec4}

We turn to the chiral anomaly for massive Dirac fermions in (1+1) dimensions using the van~Vleck high-frequency expansion.  We present our novel results for the chiral current and the pseudo-scalar condensate for massive Dirac fermions
in the high-frequency limit (with/without spatial inhomogeneity) in which
$T^{-1} \gg m,\ \sqrt{|eE|},\ L^{-1}$ is satisfied.  
We observe that the mass effects are suppressed strongly in the high-frequency limit.  Accordingly, the chirality production is not exponentially suppressed, which makes a contrast to the low-frequency limit.  Also, the chiral anomaly is saturated by the chirality production, and the pseudo-scalar condensate is subleading.  This is opposite to what we have seen in Eq.~(\ref{eq:::9}) in the low-frequency limit.  We also demonstrate that the chiral anomaly can be reproduced precisely with the low-order truncation in the van~Vleck high-frequency expansion in a way similar to the previous case with Weyl fermions. 
We here again emphasize that the chiral anomaly relation does not tell us about the separate contributions from the chiral current and the pseudo-scalar condensate, for which we need explicit calculations, and that the chiral current has more information than the chiral anomaly relation in the sense that it is more directly related to the observables of chiral physics, e.g., chiral magnetic effect.  Therefore, the results presented here is not a mere reproduction or confirmation of the anomaly relation but does have physics significance beyond it.

\subsection{Dirac equation and Hamiltonian}

Throughout this section, we employ the axial gauge,
$A^1 = 0$,
and assume that the zero mode of $A^0$ is vanishing, i.e.,
$\tilde{A}^0_0 = 0$,
for simplicity.
For the application of the Floquet theory, the gauge background is assumed to be periodic in time as
\begin{align}
	A^0(t+T,x) = A^0(t,x) \,.
\end{align}
We assign the power counting for the electric field as $E={\mathcal O}(\varepsilon^0)$ and thus $A^0 = -\int^{x} {\rm d}x' E(t,x') = {\mathcal O}(\varepsilon^0)$.
The typical length scale of the electric field $L$ should be large enough to satisfy $T/L \ll 1$, but the concrete shape can be arbitrary.
This condition would give us a convergent result in the van~Vleck expansion.
If we need the expansion for $T/L \gg 1$, we may reiterate the Floquet analysis in the spatial direction, as discussed previously in Sec.~\ref{sec3.2}.  

We adopt the following representation for $\gamma^\mu$, i.e.,
 \begin{align}
    \gamma^0 \coloneqq \sigma_x = \begin{pmatrix} 0 & 1 \\ 1 & 0 \end{pmatrix} \,,\qquad
    \gamma^1 \coloneqq -{\rm i}\sigma_y = \begin{pmatrix} 0 & -1 \\ 1 & 0 \end{pmatrix} \,,\qquad
    \gamma_5 
        \coloneqq \gamma^0 \gamma^1
        = \sigma_z = \begin{pmatrix} 1 & 0 \\ 0 & -1 \end{pmatrix} \,. \label{eq1}
\end{align}
From the Dirac equation
$[ {\rm i}\gamma^\mu(\partial_\mu + {\rm i}eA_\mu) - m ] \psi=0$, the Hamiltonian can be identified as
\begin{align}
	{\rm i}\partial_t \psi(t,x) = H(t,x) \psi(t,x)\,,\qquad
    H = \sum_{l=-\infty}^{+\infty} {\rm e}^{{\rm i}\frac{2\pi l}{T}t} \, \tilde{H}_l \label{o22}
\end{align}
with
\begin{align}
    \tilde{H}_0 = -{\rm i}\partial_x \sigma_z + m\sigma_x\,,\qquad
    \tilde{H}_{l\neq 0} = e\tilde{A}_l^0\,.  \label{eq4-7}
\end{align}
Note that we can decompose the Dirac field $\psi$ into the right- and the left-handed components as $\psi =  {}^t(\chi_{\rm R}, \chi_{\rm L})$ and confirm that the Dirac equation~(\ref{o22}) reduces to the Weyl equation~(\ref{v1}) in the $m\to 0$ limit.

\subsection{Fourth order in the van~Vleck expansion in the $A^1=0$ gauge} \label{sec4.2}

The massive Dirac equation~(\ref{o22}) is not analytically solvable even in (1+1) dimensions unlike the Weyl equation~(\ref{v1}).  Therefore, we need to truncate the van~Vleck expansion at a certain order.  Later we shall demonstrate that the mass effect in the chiral anomaly relation appears from the ${\mathcal O}(\varepsilon^{2})$ terms.  As checked in the Weyl fermion case, we should continue the van~Vleck expansion until the $(n+2)$-th order for the purpose to recover the chiral anomaly relation up to the $n$-th order.
This is because both the point-splitting regularization and the divergence of the chiral current lower the expansion order involving $\mathcal{O}(\varepsilon^{-1})$ corrections.  Thus, we should calculate the current at least to the fourth order in the van~Vleck expansion to consider the mass effect.

In the fourth-order van~Vleck expansion, the Floquet Hamiltonian $H_{\rm F}$ for the massive Dirac fermion is given by
\begin{align}
    H_{\rm F}(x) 
		&= \tilde{H}_0(x) + \delta m(x) \sigma_x + {\mathcal O}(\varepsilon^5)  \label{v23}
\end{align}
with
\begin{align}
	\delta m(x) 
		= - 2m \sum_{l\neq0} \left(\frac{T}{2\pi{\rm i}l}\right)^4  \left| \partial_xe\tilde{A}^0_{l}(x) \right|^2 
		= - 2m \sum_{l\neq0} \left(\frac{T}{2\pi{\rm i}l}\right)^4  \left| e\tilde{E}_{l}(x) \right|^2\,.
\end{align}
For the derivation, see~\ref{appfourth}.
Accordingly, the kick operator is found as
\begin{align}
    K(t,x) 
		= \kappa_0(t,x) + \kappa_x(t,x)\sigma_x + \kappa_y(t,x)\sigma_y + \kappa_z(t,x)\sigma_z \,,
    \label{v25}
\end{align}
where, up to the corrections of $\mathcal{O}(\varepsilon^5)$,
\begin{subequations}
\begin{align}
	\kappa_0 &= \sum_{l\neq0 } {\rm e}^{{\rm i}\frac{2\pi l}{T} t } \left[ \frac{T}{2\pi {\rm i}l}  e\tilde{A}^0_l + \left(\frac{T}{2\pi {\rm i} l}\right)^3  (\partial_x^2 e\tilde{A}^0_l) \right] \,, \\
	\kappa_x &= \sum_{l\neq0 } {\rm e}^{{\rm i}\frac{2\pi l}{T} t }\left[  \left(\frac{T}{2\pi {\rm i} l}\right)^4 2m{\rm i} \left( 2(\partial_xe\tilde{A}^0_l) \partial_x + (\partial_x^2e\tilde{A}^0_l) \right) \right] \,, \label{c27b} \\
	\kappa_y &= \sum_{l\neq0 } {\rm e}^{{\rm i}\frac{2\pi l}{T} t } \left[ \left(\frac{T}{2\pi {\rm i} l}\right)^3 2 m(\partial_x e\tilde{A}^0_l)  \right] \,, \\
	\kappa_z &= \sum_{l\neq0 } {\rm e}^{{\rm i}\frac{2\pi l}{T} t }\left[ - \left( \frac{T}{2\pi {\rm i}l}\right)^2 ( \partial_x e\tilde{A}^0_l) + \left(\frac{T}{2\pi {\rm i} l}\right)^4 \left( - (\partial_x^3e\tilde{A}^0_l) + 4m^2(\partial_xe\tilde{A}^0_l) \right) \right] \,.   \end{align}
\end{subequations}
It is worthwhile noting that $\delta m$ is negative definite and hence it effectively reduces the Dirac mass $m \to m + \delta m < m$.  This makes a contrast with the Volkov state under a plane-wave background in (3+1) dimensions, which predicts a positive mass shift \cite{Wolkow:1935zz}.

We can construct a wavefunction $\psi$ by using the Floquet Hamiltonian $H_{\rm F}$ in Eq.~(\ref{v23}) and the kick operator $K$ in Eq.~(\ref{v25}) as we did for the massless case.  Namely, we introduce,
\begin{align}
	\varphi(x) \coloneqq {\rm e}^{+{\rm i}K(0,x)}\psi(0,x) \,,
\end{align}
and take $\varphi$ to be an eigenfunction of the Floquet Hamiltonian~(\ref{v23}) satisfying
\begin{align}
	\lambda \varphi(x) = H_{\rm F}(x) \varphi(x) \,.
        \label{eqv22}
\end{align}
We can readily solve the eigenvalue equation (\ref{eqv22}) by employing the perturbation theory with respect to $\delta m$.  Including the leading correction of $\delta m = {\mathcal O}(\varepsilon^4)$, we find that the energy eigenvalue $\lambda$ is given by
\begin{align}
	\lambda = \lambda_\pm = \pm \left( \omega_p + \frac{m}{\omega_p} \overline{\delta m}(0) \right) + {\mathcal O}(\varepsilon^5) \,,
\end{align}
where $\omega_p\coloneqq\sqrt{m^2+p^2}$ with the kinetic momentum $p$, and
\begin{align}
	\overline{\delta m}(p) \coloneqq \int^{+V/2}_{-V/2} \frac{{\rm d}x}{V} {\rm e}^{-{\rm i}px} \delta m(x) 
\end{align}
with $V$ the total length [which equals the spatial volume in (1+1) dimensions] of the system
and we take $V\to\infty$ in the thermodynamic limit.
Dropping $\mathcal{O}(\varepsilon^5)$ corrections,
we can write down the corresponding $\varphi = \varphi_{p,\pm}$ as
\begin{subequations} \label{eqq77}
\begin{align}
	\varphi_{p,+} 
		&= u_p \frac{{\rm e}^{+{\rm i}px}}{\sqrt{2\pi}} + \frac{V}{2\pi} \int^{+\infty}_{-\infty} {\rm d}p' \overline{\delta m}(p'-p)\left[ \frac{ \left( u^\dagger_{p'} \sigma_x u_{p} \right) \left( 1 - \frac{2\pi}{V}\delta(p-p') \right) }{\omega_p-\omega_{p'}} u_{p'} + \frac{ v^\dagger_{p'} \sigma_x u_{p} }{\omega_p+\omega_{p'}} v_{p'}  \right] \frac{{\rm e}^{+{\rm i}p'x}}{\sqrt{2\pi}} \,, \\
	\varphi_{p,-} 
		&= v_p \frac{{\rm e}^{+{\rm i}px}}{\sqrt{2\pi}} - \frac{V}{2\pi} \int^{+\infty}_{-\infty} {\rm d}p' \overline{\delta m}(p'-p) \left[ \frac{u^\dagger_{p'} \sigma_x v_{p} }{\omega_p+\omega_{p'}} u_{p'}   +   \frac{ \left( v^\dagger_{p'} \sigma_x v_{p} \right) \left( 1 - \frac{2\pi}{V}\delta(p-p') \right) }{\omega_p-\omega_{p'}} v_{p'} \right] \frac{{\rm e}^{+{\rm i}p'x}}{\sqrt{2\pi}}  \,,
\end{align}
\end{subequations}
where $u_{p}$ and $v_{p}$ are the Dirac spinors, i.e.,
\begin{align}
	u_p = \frac{1}{\sqrt{2}} \begin{pmatrix} \sqrt{1+\frac{p}{\omega_p}} \\ \sqrt{1-\frac{p}{\omega_p}} \end{pmatrix} \,, \qquad 
	v_p = \frac{1}{\sqrt{2}} \begin{pmatrix} \sqrt{1-\frac{p}{\omega_p}} \\ -\sqrt{1+\frac{p}{\omega_p}}  \end{pmatrix} \,. \label{eq:spinor}
\end{align}
Note that our normalization convention gives $u_p^\dagger u_p = v_p^\dagger v_p=1$,
$u_p^\dagger v_p = 0$,
so that $\int{\rm d}x\,\varphi^\dagger_{p,\pm}\varphi_{p',\pm} = \delta(p-p')$
and
$\int{\rm d}x\,\varphi^\dagger_{p,\pm}\varphi_{p',\mp} = 0$.  We can explicitly evaluate the bispinor products as
\begin{subequations}
\begin{align}
	&u^\dagger_{p'} \sigma_x u_{p} 
		= - v^\dagger_{p'} \sigma_x v_{p} 
		=  \frac{1}{2} \left[ \sqrt{\left(1+\frac{p}{\omega_p}\right)\left(1-\frac{p'}{\omega_{p'}}\right)} + \sqrt{\left(1-\frac{p}{\omega_p}\right) \left(1+\frac{p'}{\omega_{p'}}\right)} \:\right] \,, \\
	&u^\dagger_{p'} \sigma_x v_{p} 
		= v^\dagger_{p'} \sigma_x u_{p} 
		= \frac{1}{2} \left[ \sqrt{\left(1-\frac{p}{\omega_p}\right) \left(1-\frac{p'}{\omega_{p'}}\right)} - \sqrt{\left(1+\frac{p}{\omega_p}\right) \left(1+\frac{p'}{\omega_{p'}}\right)} \:\right]  \,.
\end{align}
\end{subequations}
Note that in general the mode function $\varphi_{p,\pm}$ in Eq.~(\ref{eqq77}) cannot be simplified further.  For spatially homogeneous $E(t,x)=E(t)$, the mass correction becomes diagonal in momentum space as $\overline{\delta m}(p)=\frac{2\pi}{V}\delta(p) \delta m$ and hence we can explicitly carry out the $p'$ integration to simplify them as
\begin{subequations}
\begin{align}
	\varphi_{p,+} 
		&= \left( u_p - \delta m \frac{p}{2\omega_p^2} v_{p} \right) \frac{{\rm e}^{+{\rm i}px}}{\sqrt{2\pi}} \,, \\
	\varphi_{p,-} 
		&= \left( v_p + \delta m \frac{p}{2\omega_p^2} u_{p} \right) \frac{{\rm e}^{+{\rm i}px}}{\sqrt{2\pi}} \,,
\end{align}
\end{subequations}
dropping $\mathcal{O}(\varepsilon^5)$ terms.
Now, we can compute $\psi_{p,\pm}$ at time $t$ by applying the time-translation operator $U$ as
\begin{align}
	\psi_{p,\pm}(t,x) 
		= U(0,t;x) \psi_{p,\pm}(0,x)
		= {\rm e}^{-{\rm i}K(t,x)}{\rm e}^{-{\rm i}\lambda_\pm t } \varphi_{p,\pm}(x) \,. \label{n34}
\end{align}

We can canonically quantize the Dirac field $\psi \to \hat{\psi}$ in the usual manner.  The mode-expansion form of the field operator $\hat{\psi}$ is given by
\begin{align}
	\hat{\psi}(t,x) 
		&= \int^{+\infty}_{-\infty} {\rm d}p \left[  \hat{a}_{p} \psi_{p,+}(t,x) + \hat{b}^\dagger_{-p} \psi_{p,-}(t,x) \right] \nonumber\\
		&= \int^{+\infty}_{-\infty} {\rm d}p \, {\rm e}^{-{\rm i}K(t,x)} \left[  \hat{a}_{p} \varphi_{p,+}(x) \,{\rm e}^{-{\rm i}\lambda_+ t} + \hat{b}^\dagger_{-p} \varphi_{p,-}(x) \, {\rm e}^{-{\rm i} \lambda_- t} \right] \,, \label{bb35}
\end{align}
where $\hat{a}_p$ and $\hat{b}^\dagger_{-p}$ are interpreted as an annihilation operator for the Dirac fermion and a creation operator for the Dirac anti-fermion, respectively.  The annihilation operators $\hat{a}_p$ and $\hat{b}_{p}$ satisfy the same commutation relations as Eq.~(\ref{gg42}).  The corresponding vacuum state can be defined in the same manner as Eq.~(\ref{gg43}).  Note that the momentum $p$ runs from $-\infty$ to $+\infty$ in the mode expansion~(\ref{bb35}), which is distinct from the Weyl case in Eq.~(\ref{rr41}).  For the Weyl fermions, the momentum and the energy are locked as $\lambda=p$ and hence anti-particles are negative-momentum states.  This correspondence does not hold for the massive Dirac case, in which positive and negative energy states can have either positive or negative momentum.

\subsection{Chiral current} \label{sec4.3}

We explicitly evaluate the vacuum expectation value of the chiral current $J_5^\mu$, using the fourth-order van~Vleck results obtained in Sec.~\ref{sec4.2} and properly subtracting/renormalizing UV divergences based on the point-splitting regularization procedure.  We show that the finite mass effects are suppressed strongly by the period $T$.  We also point out that $J_5^\mu$ in the high-frequency limit exhibits distinct parameter dependence compared to the low-frequency limit.  

The expectation value of the chiral current,
$J_5^\mu(x)\coloneqq\braket{\hat{\bar{\psi}}\gamma^\mu \gamma_5 \hat{\psi}(x)}$,
for a massive Dirac fermion $\psi$ is defined as
\begin{align}
	J_5^\mu(x^\nu) 
		\coloneqq \lim_{\epsilon^\nu \to 0} {\rm e}^{-{\rm i}\int^{x^\nu+\epsilon^\nu/2}_{x^\nu-\epsilon^\nu/2}{\rm d}x^\sigma eA_\sigma } \braket{0| \hat{\bar{\psi}}(x^\nu+\epsilon^\nu/2) \gamma^\mu \gamma_5 \hat{\psi}(x^\nu-\epsilon^\nu/2) |0} \,.  \label{eq83}
\end{align} 
Plugging the fourth-order van~Vleck results, we find,
\begin{align}
	J_5^\mu 
		&= \lim_{\epsilon^\nu \to 0} {\rm e}^{-{\rm i} \epsilon^\sigma eA_\sigma(x^\nu)} \int^{+\infty}_{-\infty} {\rm d}p \ {\rm e}^{-{\rm i}\omega_p\epsilon^0} \nonumber\\
			&\quad \times \left[ \varphi_{p,-}^\dagger(x+\epsilon^1/2) {\rm e}^{+{\rm i}K(x^\nu+\epsilon^\nu/2)} \gamma^0 \gamma^\mu \gamma_5 {\rm e}^{-{\rm i}K(x^\nu-\epsilon^\nu/2)} \varphi_{p,-}(x-\epsilon^1/2) \right]  + {\mathcal O}(\varepsilon^4) \,.
\end{align}
From this expression we arrive at
\begin{subequations} \label{k40}
\begin{align}
    J_5^0 &= \frac{1}{2\pi} \lim_{\epsilon^\nu \to 0}  {\rm e}^{-{\rm i} \epsilon^\nu eA_\nu}  \left[ -I_2 + 2I_1 \kappa_y + {\rm i} \epsilon^\nu  \partial_\nu \left( - I_2\kappa_0 +  I_0 \kappa_z \right) +2{\rm i}\epsilon^\nu I_1 \kappa_y \partial_\nu \kappa_0 \right] + {\mathcal O}(\varepsilon^4) \,, \\
    J_5^1 &= \frac{1}{2\pi} \lim_{\epsilon^\nu \to 0}  {\rm e}^{-{\rm i} \epsilon^\nu eA_\nu}  \left[ I_0 + {\rm i} \epsilon^\nu \partial_\nu \left( \kappa_0 I_0 - I_2\kappa_z - I_2 m\kappa_{x,p} - I_1 \kappa_{x,0} \right) \right] + {\mathcal O}(\varepsilon^4) \,,
\end{align}
\end{subequations}
where
\begin{subequations} \label{int40}
\begin{align}
	I_0(\epsilon^\nu) 
		&\coloneqq \int^{+\infty}_{-\infty} {\rm d}p\,{\rm e}^{-{\rm i}\left( \omega_p\epsilon^0+p\epsilon^1 \right)} 
		= -\frac{{\rm i}}{\sqrt{2}}\left( \frac{1}{\epsilon^+} + \frac{1}{\epsilon^-} \right) + {\mathcal O}(|\epsilon|) , \\
	I_1(\epsilon^\nu) 
		&\coloneqq \int^{+\infty}_{-\infty} {\rm d}p\,{\rm e}^{-{\rm i}\left( \omega_p\epsilon^0+p\epsilon^1 \right)} \frac{m}{\omega_p}
		= - 2m \left( \frac{1}{2}{\rm ln}\frac{m^2|\epsilon^2|}{4} + \gamma_{\rm E} \right) + {\mathcal O}(|\epsilon|) , \\
	I_2(\epsilon^\nu) 
		&\coloneqq \int^{+\infty}_{-\infty} {\rm d}p\,{\rm e}^{-{\rm i}\left( \omega_p\epsilon^0+p\epsilon^1 \right)} \frac{p}{\omega_p}
		= -\frac{{\rm i}}{\sqrt{2}}\left( \frac{1}{\epsilon^+} - \frac{1}{\epsilon^-} \right) + {\mathcal O}(|\epsilon|)
\end{align}
\end{subequations}
with $\gamma_{\rm E} \approx 0.5772$ being Euler's constant.  As in the Weyl-fermion case, $\partial_t = {\mathcal O}(\varepsilon^{-1})$ appears from the point-splitting regularization, and the expectation value becomes one order lower than the original truncation order ${\mathcal O}(\varepsilon^5) \to {\mathcal O}(\varepsilon^4)$ in the van~Vleck expansion.  Therefore, the ${\mathcal O}(\varepsilon^4)$ terms in the Floquet Hamiltonian, $H_{\rm F}$ in Eq.~(\ref{v23}), can safely be neglected during the calculation.  This means that the effects of the mass shift $\delta m = {\mathcal O}(\varepsilon^4)$ are suppressed at this order, and they can be manifest from the fifth-order van~Vleck calculation in the $A^1=0$ gauge.   Note that $\kappa_x$ in Eq.~(\ref{c27b}) contains a derivative $\partial_x$ and that its coefficient is of the order of ${\mathcal O}(\varepsilon^4)$.  Thus, $\partial_x$ can act only onto ${\mathcal O}(\varepsilon^0)$ terms, and all other terms are of higher order and should be neglected within our approximation.  In the present case, $\partial_x$ can act only onto ${\rm e}^{+{\rm i}px}$ out of $\hat{\psi}$, and hence we can replace $\partial_x$ in $\kappa_x$ with $+{\rm i}p$, that is,
\begin{align}
	\kappa_x 
		&\to p \underbrace{ \sum_{l\neq0 } {\rm e}^{{\rm i}l \frac{2\pi}{T} t } \left[-4m \left(\frac{T}{2\pi {\rm i} l}\right)^4 (\partial_xe\tilde{A}^0_l) \right] }_{\coloneqq \kappa_{x,p}} + \underbrace{ \sum_{l\neq0 } {\rm e}^{{\rm i}l \frac{2\pi}{T} t } \left[ \left(\frac{T}{2\pi {\rm i} l}\right)^4 2m{\rm i}  (\partial_x^2e\tilde{A}^0_l)  \right] }_{\coloneqq \kappa_{x,0}} \,.  
\end{align}

The chiral current $J_5^\mu$ in Eq.~(\ref{k40}) is UV divergent, which should be removed for a physically meaningful observable.  In other words, we need to make the divergent integrals $I_0, I_2 \propto 1/|\epsilon|$ and $I_1 \propto \mathop{\mathrm{ln}}|\epsilon|$ well-defined by properly subtracting the divergences.  The former divergences are constants independent of the electric field.  Hence, as we did for the Weyl-fermion case, we can simply subtract them as $I_0, I_2 \to {\mathcal O}(|\epsilon|)$, so that the chiral current is vanishing in the absence of the electric field.  The remaining logarithmic divergence goes vanishing in the absence of the electric field, and thus the same reasoning cannot be used.  To find out a reasonable subtraction scheme, we calculate the energy density in the absence of the electric field and obtain
\begin{align}
	{\mathcal E}|_{A^0=0}
        &\coloneqq \braket{  \hat{\psi}^\dagger \frac{{\rm i}}{2} \overset{\leftrightarrow}{\partial}_t \hat{\psi} } |_{A^0=0} \nonumber\\
	&= - \lim_{\epsilon^\mu \to 0} \int^{+\infty}_{-\infty} \frac{{\rm d}p}{2\pi} \; {\rm e}^{-{\rm i}\left(\epsilon^0 \omega_p + \epsilon^1 p\right)} \omega_p \nonumber\\
	&= \frac{1}{2\pi} \lim_{\epsilon^\mu\to0} \left[ 2\left( \frac{1}{(\epsilon^+)^2} + \frac{1}{(\epsilon^-)^2}  \right)  - \frac{m}{2} I_1 + {\mathcal O}(|\epsilon|^1)  \right]\,. \label{eq:gg79}
\end{align}
The vacuum state should have vanishing energy density in the absence of the electric field.  Hence, it is natural to require 
\begin{align}
	{\mathcal E}|_{A^0=0} \xrightarrow{{\text{subtraction}}} \; 0 \,.  \label{eqre}
\end{align}
This means that both of the quadratic divergences $\propto 1/(\epsilon^\pm)^2$ and the integral $I_1$ in Eq.~(\ref{eq:gg79}) should be subtracted to be zero.  This naturally fixes the subtraction scheme for the integral $I_1$ in Eq.~(\ref{k40}), i.e., $I_1 \to {\mathcal O}(|\epsilon|)$.  

After subtracting the UV divergences in $J^\mu_5 \to [J^\mu_5]$, we find
\begin{subequations} \label{eq-240}
\begin{align}
	[J^0_5]
		&= \frac{1}{\pi} \left[ - \partial_x \int^{t}{\rm d}t' eA^0 - ( \partial_x^3 - 2m^2 \partial_x) \int^{t}{\rm d}t'\int^{t'}{\rm d}t''\int^{t''}{\rm d}t''' eA^0  \right] + {\mathcal O}(\varepsilon^4) \nonumber\\
		&= \frac{1}{\pi} \left[ \int^{t}{\rm d}t' eE + ( \partial_x^2 - 2m^2 ) \int^{t}{\rm d}t'\int^{t'}{\rm d}t''\int^{t''}{\rm d}t''' eE  \right] + {\mathcal O}(\varepsilon^4) \,, \\
	[J^1_5]
		&= \frac{1}{\pi}  \partial_x^2 \int^{t}{\rm d}t'\int^{t'}{\rm d}t'' eA^0 + {\mathcal O}(\varepsilon^4)
		= -\frac{1}{\pi} \partial_x \int^{t}{\rm d}t'\int^{t'}{\rm d}t'' eE + {\mathcal O}(\varepsilon^4)  \,.  
\end{align}
\end{subequations}
Compared to the slow-frequency result in Eq.~(\ref{eq:uu-88}) with the exponential suppression by $m^2/eE$, the high-frequency result in Eq.~(\ref{eq-240}) predicts more abundant chirality production as seen in $[J^0_5]$.  Therefore, a sizable number of chirality can be produced from the vacuum when the frequency is high enough, even if the electric field does not exceed the mass threshold.  The mass $m$ appears as an ${\mathcal O}(\varepsilon^3)$ effect in $[J^0_5]$ and the spatial component, $[J^1_5]$, has no mass dependence at this order.  All the other terms are consistent with Eq.~(\ref{rr48}) in the Weyl-fermion case.  This implies that mass effects are minor contributions in the high-frequency limit.  To see the relationship between the results in the massive Dirac and the Weyl cases, it is convenient to decompose $[J^\mu_5]$ into the right-handed and the left-handed contributions [see Eq.~(\ref{ff44})].  Namely, 
\begin{align}
	J_5^\mu = J^\mu_{\rm R} - J^\mu_{\rm L} = \begin{pmatrix} N_{\rm R} - N_{\rm L} \\ N_{\rm R} + N_{\rm L} \end{pmatrix} \,,
\end{align}
which yields
\begin{subequations}
\begin{align}
	[N_{\rm R}]
		&= \frac{[J^0_5] + [J^1_5]}{2} \nonumber\\
		&= \frac{1}{2\pi} \left[ +\int^{t}{\rm d}t' eE - \partial_x \int^{t}\!{\rm d}t' \!\int^{t'}\!{\rm d}t'' eE + ( \partial_x^2 - 2m^2) \int^{t}\!{\rm d}t'\!\int^{t'}\!{\rm d}t''\!\int^{t''}\!\!{\rm d}t''' eE  \right] + {\mathcal O}(\varepsilon^4)
 ,\\
	[N_{\rm L}]
		&= \frac{-[J^0_5] + [J^1_5]}{2} \nonumber\\
		&= \frac{1}{2\pi} \left[ - \int^{t}{\rm d}t' eE - \partial_x \int^{t}\!{\rm d}t' \!\int^{t'}\!{\rm d}t'' eE - ( \partial_x^2 - 2m^2) \int^{t}\!{\rm d}t'\!\int^{t'}\!{\rm d}t''\!\int^{t''}\!\!{\rm d}t''' eE  \right] + {\mathcal O}(\varepsilon^4) , 
\end{align}
\end{subequations}
which in the limit of $m\to 0$ agree with the Weyl results in Eq.~(\ref{rr48}).  

Incidentally, we can confirm that Eq.~(\ref{eq-240}) is compatible with gauge invariance, as it should be.  The charge current $J^\mu$ is related to the chiral current $J^\mu_5$ in (1+1) dimensions as
\begin{align}
	J^\mu 
		\coloneqq \braket{\hat{\bar{\psi}} \gamma^\mu \hat{\psi}} 
		= J_{\rm R}^\mu + J_{\rm L}^\mu
            = \begin{pmatrix} J_5^1 \\[4pt] J_5^0 \end{pmatrix} \,.
\end{align}
Thus, we can explicitly compute the divergence of the (subtracted) charge current $\partial_\mu [J^\mu]$ with Eq.~(\ref{eq-240}) as
\begin{align}
	\partial_\mu [J^\mu] = \partial_t [J^1_5] + \partial_x [J^0_5] = 0 \,,
\end{align}
up to the accuracy of our calculation, i.e., ${\mathcal O}(\varepsilon^3)$ with the time derivative $\partial_t \sim {\mathcal O}(\varepsilon^{-1})$ taken into account.  The charge current $[J^\mu]$ is therefore conserved, which is the manifestation of gauge invariance.

\subsection{Pseudo-scalar condensate}

Now we turn to evaluate the pseudo-scalar condensate $P$ defined as
\begin{align}
	P
		\coloneqq \braket{\hat{\bar{\psi}} {\rm i} \gamma_5 \hat{\psi}} 
		\coloneqq \lim_{\epsilon^\mu \to 0} {\rm e}^{-{\rm i}\int^{x^\mu+\epsilon^\mu/2}_{x^\mu-\epsilon^\mu/2}{\rm d}x^\nu eA_\nu } \braket{0| \hat{\bar{\psi}}(x^\mu+\epsilon^\mu/2) {\rm i} \gamma_5 \hat{\psi}(x^\mu-\epsilon^\mu/2) |0} \,. \label{eq--97}
\end{align} 
Using the fourth-order van~Vleck results, we find:
\begin{align}
	P 
		&= \lim_{\epsilon^\mu \to 0} {\rm e}^{-{\rm i} \epsilon^\nu eA_\nu(x^\mu)} \int^{+\infty}_{-\infty} {\rm d}p {\rm e}^{-{\rm i}\omega_p\epsilon^0}\notag\\
  &\quad\times\left[ \varphi_{p,-}^\dagger(x+\epsilon^1/2) {\rm e}^{+{\rm i}K(x^\mu+\epsilon^\mu/2)} {\rm i}\gamma^0 \gamma_5 {\rm e}^{-{\rm i}K(x^\mu-\epsilon^\mu/2)} \varphi_{p,-}(x-\epsilon^1/2) \right] + {\mathcal O}(\varepsilon^4) \nonumber\\
		&= \frac{1}{2\pi} \lim_{\epsilon^\mu \to 0}  {\rm e}^{-{\rm i} \epsilon^\nu eA_\nu}  \left[ - 2I_1 \kappa_z - 2{\rm i} \epsilon^\nu I_1 \kappa_z (\partial_\nu \kappa_0) + {\rm i}\epsilon^\nu (\partial_\nu \kappa_y) I_0 \right] + {\mathcal O}(\varepsilon^4) \,,
\end{align}
which is UV divergent.  We remove the UV divergence in the same manner as the chiral current $J^\mu_5 \to [J^\mu_5]$ [see Eq.~(\ref{eq-240}) above], which yields
\begin{align}
	[P]
	= \frac{m}{\pi} \partial_x \int^{t} \!{\rm d}t'\int^{t'} \!\!{\rm d}t'' eA^0 + {\mathcal O}(\varepsilon^4) 
	= -\frac{m}{\pi} \int^{t} \!{\rm d}t'\int^{t'}\!\!{\rm d}t'' eE + {\mathcal O}(\varepsilon^4) .  \label{y55}
\end{align}
This high-frequency result in Eq.~(\ref{y55}) has distinct parameter dependence compared to Eq.~(\ref{eq:uu-88}) in the slow limit.  Notably, Eq.~(\ref{y55}) is suppressed by the factor of $T^2$ and hence the pseudo-scalar condensate $[P]$ goes vanishing in the high-frequency limit.  This makes a sharp contrast to Eq.~(\ref{eq:uu-88}) in the slow-frequency limit, where the pseudo-scalar condensate $[P]$ takes a finite value and eventually gives the dominant contribution to the chiral anomaly.

\subsection{Chiral anomaly}

We shall discuss the massive chiral anomaly relation with the van~Vleck high-frequency expansion.  Taking the total derivative of the chiral current in Eq.~(\ref{eq-240}), we find: 
\begin{align}
    \underbrace{ \frac{eE}{\pi}}_{{\rm topological\ charge\ density}} = \underbrace{ \frac{1}{\pi}\left[ eE - 2m^2 \int^{t}\!{\rm d}t'\!\int^{t'}\!\!{\rm d}t'' eE   + {\mathcal O}(\varepsilon^3) \right]  }_{=\,\partial_\mu [J_5^\mu] \text{: chirality\ production}} + \underbrace{ \frac{1}{\pi}\left[ +2m^2 \int^{t}\!{\rm d}t'\!\int^{t'}\!\!{\rm d}t'' eE   + {\mathcal O}(\varepsilon^3) \right] }_{=\,-2m[P] \text{: pseudo-scalar condensate}} \,,\label{o56}
\end{align}
where we identified higher-order corrections as ${\mathcal O}(\varepsilon^3)$, not ${\mathcal O}(\varepsilon^4)$, because of the time derivative $\partial_t \sim {\mathcal O}(\varepsilon^{-1})$.  It is evident that Eq.~(\ref{o56}) reproduces the chiral anomaly relation with massive fermions.  The topological charge is saturated by the contribution from the chirality production in the high-frequency limit, which is qualitatively different from the slow-frequency limit as seen from Eq.~(\ref{eq:::9}) that implies that the anomaly is dominated by the pseudo-scalar condensate.  Intuitively, the mass effects are negligible in the high-frequency limit because the typical energy scale in the system is overridden by the frequency.
Hence, the existence of the mass gap should be compared with not the field strength but the frequency and would become negligible after all.  Thus, the frequency makes the massive Dirac system rather behave like a massless system, and the pseudo-scalar condensate turns out to be a minor contribution to the anomaly relation.  Equation~(\ref{o56}) also indicates that the information about the topological charge is encoded only in the low-order terms from the high-frequency expansion, as we have observed in the Weyl-fermion case.  As a matter of fact, the anomaly term $eE/\pi$ originates from the lowest order term in Eq.~(\ref{eq-240}) for $[J^0_5]$ and higher-order terms in the right-hand side of Eq.~(\ref{o56}) cancel out with each other.

\section{Standard perturbation theory} 

It is instructive to see how our Floquet treatment of the chiral anomaly is related to the standard perturbation theory with respect to the small coupling in $eA$.  We here demonstrate explicitly that two expansions are equivalent at the leading order in $eA$.  Albeit equivalent, an advantage of the Floquet treatment is that the expansion parameter is the inverse frequency $\varepsilon \sim 1/T$, not $eA$, and hence the Floquet-based treatment can be technically advantageous to look into non-linear effects in terms of $eA$.  Indeed, the Floquet wavefunction, $\psi^{\rm Floquet}_{p,\pm}$ in Eq.~(\ref{n34}), has non-linear dependence on $eA$ through, e.g., the exponential factor ${\rm e}^{-{\rm i}K}$, which is expanded in the standard perturbative treatment, as we shall see below.  Conversely, the standard perturbation theory would be useful to see some non-linear corrections in terms of $\varepsilon \sim 1/T$ for a certain fixed order in $eA$.  In general, both the chiral current and the pseudo-scalar condensate may have non-linear dependence on $eA$, e.g., Eq.~(\ref{eq:::9}) for slow electric fields clearly exhibits non-linearity.  Our Floquet results, Eqs.~(\ref{eq-240}) and (\ref{y55}), can be regarded as a theoretical verification that the chiral current and the pseudo-scalar condensate for fast electric fields do not receive any non-linear corrections in $eA$ up to ${\mathcal O}(\varepsilon^4)$.  It is not easy to prove this statement within the framework of the standard perturbation theory.

Let us compute the wavefunction $\psi^{\rm pert}_{p,\pm}$ based on the standard perturbation theory at the leading order in $eA$ and it is interesting to compare it with the Floquet wavefunction, $\psi^{\rm Floquet}_{p,\pm}$ in Eq.~(\ref{n34}).  Introducing the retarded Green's function, $S_{\rm R}$, such that
\begin{align}
    [{\rm i}\slashed{\partial}-m]S_{\rm R}(t-t',x-x')=\delta(t-t')\delta(x-x')\ \ {\rm and} \ \ 
    S_{\rm R}(t-t'<0)=0 \,, 
\end{align}
we can readily solve the Dirac equation~(\ref{o22}) as  
\begin{align}
	\psi^{\rm pert}_{p,\pm}(t,x)
		&= \begin{pmatrix} u_p {\rm e}^{-{\rm i}\omega_pt} \\  v_p {\rm e}^{+{\rm i}\omega_pt} \end{pmatrix} \frac{{\rm e}^{+{\rm i}px}}{\sqrt{2\pi}} + \int^{+\infty} {\rm d}t'{\rm d}x' S_{\rm R}(t-t',x-x') e\slashed{A}(t',x') \begin{pmatrix} u_p {\rm e}^{-{\rm i}\omega_pt'} \\  v_p {\rm e}^{+{\rm i}\omega_pt'} \end{pmatrix} \frac{{\rm e}^{+{\rm i}px'}}{\sqrt{2\pi}}   + {\mathcal O}(|eA|^2) \,. \label{eq-100}
\end{align}
By using the expansion of the translation operator, $eA(t',x')=\sum_{n=0}^\infty \frac{(x'-x)^n}{n!} \partial_x^n eA(t',x)$, we can carry out the $x'$ integration, which yields $\partial_p^n \delta(p)$.  Also, we note that an explicit form of the retarded Green's function reads:
\begin{align}
	& S_{\rm R}(t-t';x-x') \notag\\
		&= -{\rm i}\Theta(t-t') \int {\rm d}p \frac{{\rm e}^{+{\rm i}p(x-x')}}{2\pi}\left[ u_p\bar{u}_{p} {\rm e}^{-{\rm i}\omega_p(t-t')} + v_p\bar{v}_{p} {\rm e}^{+{\rm i}\omega_p(t-t')} \right] \notag\\
		&= -{\rm i}\Theta(t-t') \int {\rm d}p \frac{{\rm e}^{+{\rm i}p(x-x')}}{2\pi} \sum_{n=0}^\infty ({\rm i}\omega_p)^{2n} \left[ \sigma_x  \frac{(t-t')^{2n}}{(2n)!} + (p\sigma_y-{\rm i}m)  \frac{(t-t')^{2n+1}}{(2n+1)!} \right]  \,,
\end{align}
where $\Theta(t-t')$ is the Heaviside step function.
Using the Dirac equations, $+{\rm i}\omega_p \sigma_x u_p = ( p \sigma_y + {\rm i}m) u_p$ and $-{\rm i}\omega_p \sigma_x v_p = ( p \sigma_y + {\rm i}m) v_p$, we can cast the perturbative solution~(\ref{eq-100}) into the form of
\begin{align}
	&\psi^{\rm pert}_{p,\pm}(t,x)
	= \Biggl[ 1  - {\rm i} \sum_{n=0}^\infty \sum_{n'=0}^{\infty} \sum_{n''=0}^\infty  \frac{1}{n'! n''!} \int^t {\rm d}t' \biggl\{ (-{\rm i}\partial_{p})^{n'} ({\rm i}\omega_p)^{2n} \biggl( \sigma_x  \frac{(t-t')^{2n+n''}}{(2n)!} \notag\\
       & \quad + (p\sigma_y-{\rm i}m)  \frac{(t-t')^{2n+1+n''}}{(2n+1)!} \biggr) \biggr\}
	(\partial_x^{n'} e\slashed{A}(t',x)) ( p {\rm i}\sigma_z + {\rm i}m \sigma_x)^{n''} \Biggl] \begin{pmatrix} u_p {\rm e}^{-{\rm i}\omega_pt} \\  v_p {\rm e}^{+{\rm i}\omega_pt} \end{pmatrix} \frac{{\rm e}^{+{\rm i}px}}{\sqrt{2\pi}} + {\mathcal O}(|eA|^2)\,.  
\end{align}
Note that terms with $n'>2n+1$ are vanishing because the largest power of $p$ inside the curly brackets appears from $\omega_p^{2n}p \sim p^{2n+1}$.  Using the identity $\int^t {\rm d}t_{n} \int^{t_n} {\rm d}t_{n-1} \cdots \int^{t_2} {\rm d}t_{1} f(t_1) = \int^t {\rm d}t' \frac{(t-t')^{n-1}}{(n-1)!} f(t')$ and introducing $\ell \coloneqq 2n+n''$, we find:
\begin{align}
	&\psi^{\rm pert}_{p,\pm}(t,x)
	= \Biggl[ 1  - {\rm i} \sum_{\ell=0}^\infty \sum_{n'=0}^{\infty} \sum_{n''=\ell,\ell-2,\cdots}  \frac{1}{n'! n''!} \biggl\{ (-{\rm i}\partial_{p})^{n'} ({\rm i}\omega_p)^{\ell-n''} \biggl( \sigma_x  \frac{\ell!}{(\ell-n'')!} \notag\\
 &\quad \times\int^t {\rm d}t_{\ell+1}\! \int^{t_{\ell+1}}\!\!{\rm d}t_{\ell} \cdots \int^{t_{2}}\!{\rm d}t_{1} + (p\sigma_y-{\rm i}m)  \frac{(\ell+1)!}{(\ell+1-n'')!} \int^t {\rm d}t_{\ell+2} \!\int^{t_{\ell+2}}\!\!{\rm d}t_{\ell+1} \cdots \int^{t_{2}}\!{\rm d}t_{1} \biggr) \biggr\} \nonumber\\
		       &\quad \times (\partial_x^{n'} e\slashed{A}(t_1,x)) ( p {\rm i}\sigma_z + {\rm i}m \sigma_x)^{n''}  \Biggl] \begin{pmatrix} u_p {\rm e}^{-{\rm i}\omega_pt} \\  v_p {\rm e}^{+{\rm i}\omega_pt} \end{pmatrix} \frac{{\rm e}^{+{\rm i}px}}{\sqrt{2\pi}} + {\mathcal O}(|eA|^2) \,.  \label{eq105}
\end{align}
It is tedious but is straightforward to evaluate Eq.~(\ref{eq105}) order-by-order in $\ell$.  Identifying $\int {\rm d}t \sim {\mathcal O}(\varepsilon)$, we see that the computation up to $\ell=4$ suffices for the ${\mathcal O}(\varepsilon^5)$ result.  The final expression is
\begin{align}
    \psi^{\rm pert}_{p,\pm}(t,x) = \left[ 1 - {\rm i}K(t,x) \right] \begin{pmatrix} u_p {\rm e}^{-{\rm i}\omega_pt} \\  v_p {\rm e}^{+{\rm i}\omega_pt} \end{pmatrix} \frac{{\rm e}^{+{\rm i}px}}{\sqrt{2\pi}} + {\mathcal O}((eA)^2, \varepsilon^5)
\end{align}
with $K$ given by Eq.~(\ref{v25}).  This is in perfect agreement with the Floquet wavefunction,
$\psi^{\rm Floquet}_{p,\pm}$ in Eq.~(\ref{n34}), after expanding ${\rm e}^{-{\rm i}K} = 1 - {\rm i}K + {\mathcal O}((eA)^2)$ and dropping the mass shift $\delta m \sim {\mathcal O}((eA)^2)$.

\section{Summary and discussion}

We have investigated the chiral anomaly for the Weyl and the massive Dirac fermions in (1+1) dimensions using the Floquet theory for the first time on the basis of the van~Vleck high-frequency expansion.

As a first study of the application of the Floquet theory to the chiral anomaly, we first considered the Weyl fermion (see Sec.~\ref{sec3}), which is exactly solvable and can be discussed in detail by analytical means.  To understand how the chiral anomaly can be understood in the Floquet language and also the mathematical structure behind there, we have computed the Floquet Hamiltonian, $H_{\rm F}$, and the kick operator, $K$, to arbitrary orders; see Eqs.~(\ref{oo24}) and (\ref{ww30}), respectively, for the final results.  We have shown that only $K$ receives corrections from the periodic driving, and therefore nontrivial physics including the chiral anomaly can arise from the kick operator $K$.  Using these all-order expressions and employing the point-splitting regularization scheme to compute the vacuum expectation values in a gauge-invariant manner, we have exactly reproduced the chiral anomaly relation and found that only the first two terms in the high-frequency expansion contain the full information about the topological charge and that the higher-order terms with $n>2$ cancel out with each other.  This implies an intriguing possibility that the information on the chiral anomaly could be dominantly encoded in the low-order terms generally in the high-frequency expansion.  We have also discussed the convergence condition for the van~Vleck expansion, noticing that it converges only for $T/L<1$, with $T$ and $L$ being the period and the typical spatial length of the applied electric field, respectively, and that we may investigate even the apparently nonconvergent regime $T/L> 1$ via resummation and/or applying the Floquet theory to the most rapidly oscillatory direction.

We then turn to a more non-trivial and realistic situation of the massive Dirac fermion (see Sec.~\ref{sec4}).  We have employed the fourth-order van~Vleck expansion in the axial $A^1=0$ gauge and appropriately subtracted/renormalized the UV divergence to obtain analytical expressions for the chiral current (\ref{eq-240}) and the pseudo-scalar condensate (\ref{y55}).  To the best of our knowledge, this is the very first study presenting the analytical expressions for the chiral current and the pseudo-scalar condensate (for the massive Dirac system) in the high-frequency limit, with or without spatial inhomogeneity.  We have found that the mass effects are suppressed by powers of the period (or the inverse frequency).  Accordingly, the chirality production is enhanced compared to the slow-frequency limit in which the chirality production is exponentially suppressed.  Also, we have shown that the chiral anomaly relation can be reproduced by the low-order terms in the high-frequency expansion, as implied from the Weyl fermion result.  Unlike the slow-frequency limit, the anomaly relation is saturated by the contribution from the chirality production.  Therefore, the physical composition of the topological charge changes, depending on the frequency of the applied electric field.  Moreover, we have demonstrated that the Dirac mass effect is in practice reduced due to the periodic driving in the high-frequency limit.  As a technical outcome, we have revealed that, similarly to the Weyl case, the information of the chiral anomaly is stored in the kick operator, not in the Floquet Hamiltonian, in the high-frequency limit.  This is theoretically important, as it means that the kick operator is the essence of the chiral anomaly in driven systems and that the chiral anomaly can never be accessed simply by analyzing the Floquet Hamiltonian, which is often the calculation target of the literature of the Floquet analysis.  

There are several interesting implications/applications of our results.  The first example is the application to condensed-matter systems: (1+1)-dimensional systems including fermionic degrees of freedom could be realized with actual experiments/materials by utilizing, e.g., synthetic gauge fields and quantum wires.  Loosely speaking, the chiral current in (1+1) dimensions amounts to the number difference between right- and left-moving fermions, and hence our predictions may be directly tested by counting the number difference at the right- and the left-edge of a (1+1)-dimensional material under consideration.  As far as we are aware, the massive chiral anomaly relation, in particular, the physical composition of the topological charge, has not been tested in measurable condensed-matter setups.  Also, they may provide suggestive information to high-energy phenomena such as the chiral magnetic effect in relativistic heavy-ion collision experiments.  In this context, the mass effect is still an open issue and it could modify the actual observables.  As chirality in (1+1) dimensions is equivalent to the number of right- and left-moving fermions, our prediction on the chirality production can be tested by counting the number difference at the right- and left-edges of a (1+1)-dimensional material under consideration.  Our results indicate that the chirality production (i.e., the number difference at the edges) will be enhanced by making the fields faster and faster and shows power dependence on the field strength, which is in contrast to the strong exponential suppression for the slow electric fields; see Eqs.~(\ref{eq:::9}) and (\ref{o56}).   Our formula (\ref{o56}) also quantifies how much the chirality production deviates from the topological charge density due to the pseudo-scalar contribution.  This can be used to remove the contamination by the residual-mass effect in experiments with Weyl semi-metals, as the pseudo-scalar contribution cannot be measured directly in experiments and therefore the theory input, i.e., Eq.~(\ref{o56}), is needed.  Conversely, it is possible to determine the value of the pseudo-scalar condensate indirectly in a given periodic system by measuring the chirality production with Eq.~(\ref{o56}) and a given field strength $eE$.  To the best of our knowledge, the value of the pseudo-scalar condensate has never been measured in experiments, and therefore it is of interest.  Our result (\ref{o56}) predicts that the pseudo-scalar contribution vanishes in the high-frequency limit as $\propto T^{2} \to 0$.  The condensed-matter application discussed above may provide suggestive information to high-energy phenomena such as the chiral-magnetic effect in relativistic heavy-ion collision experiments, where the mass effect is still an open issue and might affect the actual observables.  Another interesting implication of our results is a possible negative mass shift due to high-frequency electric fields.  There have been proposals to measure the effective mass in intense-laser facilities such as the Extreme Light Infrastructure (ELI).  The Volkov state is the widely-used description for the mass shift, which predicts that the mass shift is definitely positive~\cite{Wolkow:1935zz}.  Our results would suggest that the mass shift is not necessarily positive definite, depending on profiles and/or parameters of the applied field.  Unfortunately, it is still challenging to produce such high-frequent electric fields with intense lasers within the current technology [where the frequency is roughly around ${\mathcal O}(1\,\text{eV - } 1\,\text{keV})$].  The reachable frequency is much smaller compared to the electron mass scale~\cite{DiPiazza:2011tq, Fedotov:2022ely}) and hence our predictions may be unable to be confirmed right away.  Nevertheless, it is an important theoretical step to clarify the conditions when/how the effective mass can become different from that in the Volkov state for future developments (see also Ref.~\cite{PhysRevLett.109.100402}).  Meanwhile, the reduction of mass by high-frequency fields implies that the Schwinger mechanism by slow electric fields should be enhanced in a more straightforward manner than the dynamically assisted Schwinger mechanism~\cite{Schutzhold:2008pz, DiPiazza:2009py, Dunne:2009gi, Copinger:2016llk}, when our Floquet vacuum is exposed to both fast and slow electric fields.  In the literature, there are not many studies on the dynamical assistance with such high-frequency fields exceeding the mass scale~\cite{Taya:2018eng, Huang:2019szw, Taya:2020bcd, Villalba-Chavez:2019jqp}, and our Floquet approach may found a novel theoretical ground for this problem.

Finally, we mention that our (1+1)-dimensional Floquet analysis is a first step toward attacking the chiral anomaly in more complicated (1+3)-dimensional [or general (1+$d$)-dimensional] systems under a periodic driving including the mass effects.  The massive chiral anomaly relation, including the physical composition of the topological charge, has not been fully understood particularly in the high-frequency regime, where the conventional physics picture based on the spectral flow and the Schwinger mechanism should be replaced by the expansion with the complete basis of the Floquet states.  According to our (1+1)-dimensional results, the Floquet approach provides a powerful framework to discuss the chiral anomaly, as complementary to the standard perturbation.  Since the essential feature of the chiral anomaly is captured even in the present (1+1)-dimensional consideration, we may naturally expect that the Floquet approach should extend well to the generic (1+$d$)-dimensional problems.  This deserves future investigations.

\section*{acknowledgments}
The authors thank the domestic molecule-type workshop ``Chiral Anomaly in Periodically Driven Systems" (YITP-T-21-04) at the Yukawa Institute for Theoretical Physics in Kyoto University.  The authors also thank Takashi~Oka for enlightening discussions.  This work was supported by Japan Society for the Promotion of Science (JSPS) KAKENHI Grant Numbers 22H01216, 22H05118 (K.F.), 21H01084 (Y.H.), 21J11298 (T.S.), 22K14045 (H.T.), and by the RIKEN special postdoctoral researcher program (H.T.).  Y.H. and H.T. thank iTHEMS NEW working group for fruitful discussions.

\appendix

\section{Derivation of the fourth-order Floquet Hamiltonian and kick operator within the van~Vleck expansion in $A^1=0$ gauge} \label{appfourth}

We here explain some details of the derivation of the fourth-order van~Vleck expressions for the Floquet Hamiltonian (\ref{v23}) and the kick operator (\ref{v25}) in $A^1=0$ gauge.  
Up to the second order, the general expressions for the Floquet Hamiltonian and the kick operator are given in Eqs.~(\ref{oo16}) and (\ref{oo17}), respectively.   Just by substituting the Hamiltonian (\ref{eq4-7}) into them, we can obtain $H_{\rm F}^{(0)}, H_{\rm F}^{(1)}, H_{\rm F}^{(2)}, K^{(0)}, K^{(1)}$, and $K^{(2)}$.  It is also easy to obtain the third-order kick operator $K^{(3)}$, which can be achieved by integrating Eq.~(\ref{a182c}).  In such a manner, we can show,
\begin{align}
	\tilde{H}_{\rm F}^{(0)} = \tilde{H}_0,\ 
	\tilde{H}_{\rm F}^{(1)} = \tilde{H}_{\rm F}^{(2)} = 0,   
\end{align}
and
\begin{subequations}
\begin{align}
	\tilde{K}^{(1)}_l 
		&= \frac{T}{2\pi {\rm i} l} \tilde{H}_{l} ,  \\
	\tilde{K}^{(2)}_l 
		&= + {\rm i} \left(\frac{T}{2\pi {\rm i} l}\right)^2 \left[ \tilde{H}_l , \tilde{H}_0 \right] ,  \\
	\tilde{K}_l^{(3)} 
		&= + \left(\frac{T}{2\pi {\rm i} l}\right)^3    \left[ \tilde{H}_{0} , \left[ \tilde{H}_l , \tilde{H}_0 \right]  \right]  , 
\end{align}  
\end{subequations}
for $l\neq 0$ and $\tilde{K}^{(n)}_0 = 0$.

Below, we compute the remaining objects $H_{\rm F}^{(3)}, K^{(3)}$, and $K^{(4)}$.  The basic strategy is completely the same as what we have outlined in Sec.~\ref{sec2.2}: we solve Eq.~(\ref{xx172}) order by order in $\varepsilon$ under the van~Vleck gauge fixing condition (\ref{vv178}), i.e., $\tilde{K}^{(n)}_0 = 0 $.  At the third and fourth orders, the equations that we need to solve are
\begin{align}
	H_{\rm F}^{(3)} 
		&= + {\rm i}[K^{(3)},H] -\frac{1}{2}[K^{(2)},[K^{(1)},H]] -\frac{1}{2}[K^{(1)},[K^{(2)},H]] -\frac{{\rm i}}{6} [K^{(1)},[K^{(1)},[K^{(1)},H]]]  \nonumber\\
				&\quad  -\dot{K}^{(4)} -\frac{{\rm i}}{2}[K^{(1)},\dot{K}^{(3)}]  -\frac{{\rm i}}{2}[K^{(2)},\dot{K}^{(2)}] -\frac{{\rm i}}{2}[K^{(3)},\dot{K}^{(1)}]  \nonumber\\
				&\quad  + \frac{1}{6}[K^{(2)},[K^{(1)},\dot{K}^{(1)}]] + \frac{1}{6}[K^{(1)},[K^{(2)},\dot{K}^{(1)}]] + \frac{1}{6}[K^{(1)},[K^{(1)},\dot{K}^{(2)}]] \nonumber\\
				&\quad + \frac{{\rm i}}{24}[K^{(1)},[K^{(1)},[K^{(1)},\dot{K}^{(1)}]]] ,  \label{a-182d} 
\end{align}
and
\begin{align}
	H_{\rm F}^{(4)}
		&= +{\rm i}[K^{(4)},H] 
							-\frac{1}{2}[K^{(1)},[K^{(3)},H]] -\frac{1}{2}[K^{(2)},[K^{(2)},H]] -\frac{1}{2}[K^{(3)},[K^{(1)},H]] \nonumber\\&\quad
							- \frac{{\rm i}}{6} [K^{(1)},[K^{(1)},[K^{(2)},H]]] -\frac{{\rm i}}{6} [K^{(1)},[K^{(2)},[K^{(1)},H]]] -\frac{{\rm i}}{6} [K^{(2)},[K^{(1)},[K^{(1)},H]]]\nonumber\\&\quad
							+ \frac{1}{24}[K^{(1)},[K^{(1)},[K^{(1)},[K^{(1)},H]]]] - \dot{K}^{(5)}  \nonumber\\&\quad
							- \frac{{\rm i}}{2}[K^{(1)},\dot{K}^{(4)}] - \frac{{\rm i}}{2}[K^{(2)},\dot{K}^{(3)}] - \frac{{\rm i}}{2}[K^{(3)},\dot{K}^{(2)}] - \frac{{\rm i}}{2}[K^{(4)},\dot{K}^{(1)}] \nonumber\\&\quad
							+ \frac{1}{6}[K^{(1)},[K^{(1)},\dot{K}^{(3)}]] + \frac{1}{6}[K^{(1)},[K^{(2)},\dot{K}^{(2)}]] + \frac{1}{6}[K^{(2)},[K^{(1)},\dot{K}^{(2)}]] \nonumber\\&\quad
							+ \frac{1}{6}[K^{(1)},[K^{(3)},\dot{K}^{(1)}]] + \frac{1}{6}[K^{(2)},[K^{(2)},\dot{K}^{(1)}]] + \frac{1}{6}[K^{(3)},[K^{(1)},\dot{K}^{(1)}]] \nonumber\\&\quad
							+ \frac{{\rm i}}{24}[K^{(1)},[K^{(1)},[K^{(1)},\dot{K}^{(2)}]]] + \frac{{\rm i}}{24}[K^{(1)},[K^{(1)},[K^{(2)},\dot{K}^{(1)}]]] + \frac{{\rm i}}{24}[K^{(1)},[K^{(2)},[K^{(1)},\dot{K}^{(1)}]]] \nonumber\\&\quad
							+ \frac{{\rm i}}{24}[K^{(2)},[K^{(1)},[K^{(1)},\dot{K}^{(1)}]]] 
							- \frac{1}{120}[K^{(1)},[K^{(1)},[K^{(1)},[K^{(1)},\dot{K}^{(1)}]]]] . \label{a-182e}
\end{align}
For our Hamiltonian (\ref{eq4-7}), $\tilde{H}_l$'s ($l\neq 0$) are just scalars and thanks to this many of the commutations in Eqs.~(\ref{a-182d}) and (\ref{a-182e}) are vanishing.  After simplifying Eq.~(\ref{a-182d}), we find,
\begin{align}
	H_{\rm F}^{(3)} 
		&= + {\rm i}[K^{(3)},\tilde{H}_{0}]  - \dot{K}^{(4)}  \nonumber\\
		&= \sum_{l} {\rm e}^{{\rm i}l\frac{2\pi}{T}t} \left[+ {\rm i}[\tilde{K}^{(3)}_l,\tilde{H}_{0}](1-\delta_{l,0}) - \frac{2\pi{\rm i}l}{T}\tilde{K}^{(4)}_l(1-\delta_{l,0})  \right] , \label{gga5}
\end{align}
where 
\begin{align}
	[\tilde{K}^{(3)}_l,\tilde{H}_{0}]
	= + \left(\frac{T}{2\pi {\rm i} l}\right)^3 \left( \left\{ +{\rm i} (\partial_x^3e\tilde{A}^0_l) -4{\rm i}m^2(\partial_xe\tilde{A}^0_l) \right\} \sigma_z + \left\{ 4m(\partial_xe\tilde{A}^0_l) \partial_x + 2m(\partial_x^2e\tilde{A}^0_l) \right\} \sigma_x \right) .   
\end{align}
The zero mode and nonzero modes of the right-hand side of Eq.~(\ref{gga5}) give $H_{\rm F}^{(3)}$ and $K^{(4)}$, respectively.  Thus,
\begin{align}
	H_{\rm F}^{(3)} = 0,
\end{align}
and
\begin{align}
	\tilde{K}^{(4)}_l
	= {\rm i} \left(\frac{T}{2\pi {\rm i} l}\right)^4 \left( \left\{ +{\rm i} (\partial_x^3e\tilde{A}^0_l) -4{\rm i}m^2(\partial_xe\tilde{A}^0_l) \right\} \sigma_z + \left\{ 4m(\partial_xe\tilde{A}^0_l) \partial_x + 2m(\partial_x^2e\tilde{A}^0_l) \right\} \sigma_x \right) .
\end{align}
Notice that $\tilde{K}^{(4)}_l$ contains a derivative operator $\partial_x$ in the third term.  Next, we turn to simplify Eq.~(\ref{a-182e}) to determine $H_{\rm F}^{(4)}$.  Since $H_{\rm F}^{(4)}$ contains only the zero mode, we may neglect nonzero modes in the right-hand side of Eq.~(\ref{a-182e}).  It is straightforward to show
\begin{align}
	H_{\rm F}^{(4)}
		&= +{\rm i}\sum_{l'\neq0}[\tilde{K}_{-l'}^{(4)},\tilde{H}_{l'}] - \frac{1}{2}\sum_{l'\neq0}[\tilde{K}^{(1)}_{-l'},[\tilde{K}^{(3)}_{l'},\tilde{H}_0]] -\frac{1}{2}\sum_{l'\neq 0}[\tilde{K}^{(2)}_{-l'},[\tilde{K}^{(2)}_{l'},\tilde{H}_0]] -\frac{1}{2}\sum_{l'\neq 0}[\tilde{K}_{-l'}^{(3)},[\tilde{K}_{l'}^{(1)},\tilde{H}_0]] \nonumber\\
			&\quad  - \frac{{\rm i}}{2}\sum_{l'\neq 0} \frac{2\pi{\rm i}l'}{T} [\tilde{K}^{(1)}_{-l'},\tilde{K}^{(4)}_{l'}]  - \frac{{\rm i}}{2}\sum_{l'\neq 0}\frac{2\pi{\rm i}l'}{T}[\tilde{K}^{(2)}_{-l'},\tilde{K}^{(3)}_{l'}] - \frac{{\rm i}}{2}\sum_{l'\neq0}\frac{2\pi{\rm i}l'}{T}[\tilde{K}^{(3)}_{-l'},\tilde{K}^{(2)}_{l'}] \nonumber\\
			&\quad - \frac{{\rm i}}{2}\sum_{l'\neq0}\frac{2\pi{\rm i}l'}{T}[\tilde{K}^{(4)}_{-l'},\tilde{K}^{(1)}_{l'}] \nonumber\\
		&= \sum_{l\neq0} \left(\frac{T}{2\pi{\rm i}l}\right)^4  \frac{1}{2} [\tilde{H}_{-l},[[\tilde{H}_0,[\tilde{H}_{l},\tilde{H}_0]],\tilde{H}_0]] \nonumber\\
		&= - 2m \sum_{l\neq0} \left(\frac{T}{2\pi{\rm i}l}\right)^4  \left| \partial_xe\tilde{A}^0_{l} \right|^2 \sigma_x .
\end{align}
Collecting everything above, we arrive at the fourth-order expressions, (\ref{v23}) and (\ref{v25}), in the main text.

\bibliographystyle{elsarticle-num}
\bibliography{refs}

\end{document}